\documentclass[preprint,12pt]{elsarticle}

\usepackage{amsfonts}
\usepackage{amssymb}
\usepackage{amsmath}
\usepackage{graphicx}
\usepackage{algorithm}
\usepackage{algorithmic}
\usepackage{subfig}
\usepackage{color}
\usepackage{psfrag}
\usepackage{mathrsfs}
\usepackage{cases}
\usepackage{hyperref}

\newcommand{\argmin}{\operatornamewithlimits{argmin}}
\newcommand{\unit}[1]{\ensuremath{\mathrm{#1}}}

\begin{document}

\begin{frontmatter}

\title{Rotationally Invariant Image Representation for Viewing Direction Classification in Cryo-EM}
\author[CIMS]{Zhizhen~Zhao\corref{cor1}} 
\ead{jzhao@cims.nyu.edu}

\author[PACM]{Amit~Singer}
\ead{amits@math.princeton.edu}

\cortext[cor1]{Corresponding author. Address: Courant Institute of Mathematical Sciences, New York University, Warren Weaver Hall, 251 Mercer street, New York NY 10012 USA. Tel: +1 (443)-791-4306.}
\address[CIMS]{Courant Institute of Mathematical Sciences, New York University, Warren Weaver Hall, 251 Mercer street, New York NY 10012 USA.}

\address[PACM]{Department of Mathematics and PACM, Princeton University, Fine Hall, Washington Road, Princeton NJ 08544-1000 USA.}

\begin{abstract}We introduce a new rotationally invariant viewing angle classification method for identifying, among a large number of cryo-EM projection images, similar views without prior knowledge of the molecule. Our rotationally invariant features are based on the bispectrum. Each image is denoised and compressed using steerable principal component analysis (PCA) such that rotating an image is equivalent to phase shifting the expansion coefficients. Thus we are able to extend the theory of bispectrum of 1D periodic signals to 2D images. The randomized PCA algorithm is then used to efficiently reduce the dimensionality of the bispectrum coefficients, enabling fast computation of the similarity between any pair of images. The nearest neighbors provide an initial classification of similar viewing angles. In this way, rotational alignment is only performed for images with their nearest neighbors. The initial nearest neighbor classification and alignment are further improved by a new classification method called vector diffusion maps. Our pipeline for viewing angle classification and alignment is experimentally shown to be faster and more accurate than reference-free alignment with rotationally invariant K-means clustering, MSA/MRA 2D classification, and their modern approximations.
\end{abstract}

\begin{keyword}
Cryo-EM \sep 2D classification \sep single particle reconstruction
\end{keyword}
\end{frontmatter}
\section{Introduction}
Single particle reconstruction (SPR) from cryo-electron microscopy (EM) images is an entirely general technique for determining the 3D structures of macromolecular complexes~\cite{Frank2006, vanHeel2000, Sigworth2006, Frank2009}, which does not require crystallization or other special preparation of the complexes to be imaged. In cryo-EM, the functionally active macromolecular complexes are prepared in vitro, stalled by chemical means, and rapidly frozen by immersion into liquid ethane at liquid-nitrogen temperature. The randomly oriented and positioned macromolecular ``particles'', typically complexes 200 kDa or larger in size, are maintained at the liquid-nitrogen temperature throughout the image acquisition in the microscope. One of the challenges in SPR with cryo-EM images is the low signal to noise ratio (SNR), due to the lack of periodicity of the molecule frozen in thin vitreous ice layer. 

Because of the low SNR, it is extremely hard to visualize individual particle. To improve the resolution, a crucial step is alignment and averaging of the 2D projection images, a procedure known as ``class averaging''. Images from the same projection angles should be identified, centered, rotationally aligned and averaged to achieve a higher SNR.  Generating 2D class averages could be useful for common-lines based 3D {\it ab initio} reconstruction. They can also be used for direct observation to look for heterogeneity or discover symmetry as well as for separating particles into subgroups for additional analysis. Therefore, it is important to have fast and accurate algorithms for computing class averages. 

There are two main approaches for generating 2D class averages. IMAGIC~\citet{IMAGIC1996} uses multivariate statistical analysis (MSA) and multi-reference alignment (MRA) for 2D image classification. The MSA compresses and denoises large image data sets to achieve efficient classification using hierachical ascending classification method. The clustered images produce references for the MRA class averaging step. Since projection images can be similar up to rotation and small translations, several invariant features were proposed as a preprocessing step for viewing angle classification, for example, autocorrelation functions (ACF) and double autocorrelation function (DACF)~\citet{vanHeel1990a}. SPIDER~\citet{SPIDERprotocol} uses reference-free alignment (RFA)~\citet{Penczek1992} followed by rotationally invariant K-means clustering~\citet{Penczek1996} for 2D class averaging. Reference-free alignment tries to globally align images. The optimization method aims at finding alignment parameters of rotations and shifts for all images that minimize the sum of squared deviations from their mean (i.e., minimum variance). 

Modern software packages for SPR also include procedures for 2D class averaging. EMAN2~\citet{Tang2007} 2D class averaging method uses invariant features for initial classification. The calculation of invariants is a 2-stage process. It first computes the self correlation function (SCF)~\citet{vanHeel1992} of an image to make it translational invariant, which is followed by a polar transformation and a sequence of 1-D autocorrelations on each ring to generate rotationally invariant SCF images. The invariants are only used to bootstrap the process and the classification after this point is MSA/MRA based. The procedure for 2D class averaging in Xmipp~\citet{Xmipp3} is CL2D, which is based on the algorithm proposed by Sorzano et al~\citet{Sorzano2010}. Their algorithm for 2D multireference alignment and classification is based on a hierachical clustering approach using correntropy instead of the traditional correlation. Computing the correntropy between each image and the class reference gives classification results that are less sensitive to noise. They also proposed a new clustering criterion so as to avoid the situation that the cleaneast class ``attracts'' many experimental images even if they belong to some other classes. This modified criterion for the definition of the clusters was shown to be especially suited for images with low SNR. SPARX~\citet{SPARX} uses a 2D class averaging method called iterative stable alignment and clustering (ISAC)~\citet{ISAC}, which relies on the concepts of stability and reproducibility of clusters. Relion~\citet{Scheres2012} uses a Bayesian approach to infer parameters for a statistical model from the data. This method is used in both reference-free 2D class averaging and unsupervised 3D classification. The class averages can be deblurred and refined by using algorithms proposed in~\citet{Park2010, Park2011, Park2014}.

We notice that RFA produces significantly large errors when the images have many different views. The reason for this failure is mathematical: there does not exist an assignment of in-plane rotational angles that can align all images simultaneously. The underlying theorem is known as the hairy ball theorem, and we will elaborate on this issue in the following section. While global alignment is impossible, one can always determine the rotationally invariant distances between all pairs of images by optimally aligning each pair of them. In this way, we have to perform ${n \choose 2}$ alignments for $n$ images. This is computationally intensive and unnecessary, because most of the time is spent on aligning images from very different views. It would be more efficient to use a rotationally invariant representation for the images, then find neighboring images, and finally align and average only neighboring images.

We introduce a new rotationally invariant representation for computing the rotationally invariant distance between all pairs of cryo-EM images. Our invariant representation is based on expanding the images in a steerable basis and deriving a bispectrum for this expansion~\citet{Ponce2010, Zhao2013}. Unlike ACF, DACF and SCF, the new rotationally invariant representation maintains phase information and is complete, in the sense of uniquely specifying the original image up to an arbitrary rotation. In signal and image processing, a wide variety of invariants were devised for pattern recognition~\citet{Michaelis1995}. A common feature of most invariants is that they are lossy, in the sense that they do not uniquely specify the original signal. Among invariant features, the bispectrum and the triple-correlation function provide a lossless shift-invariant representation, and various algorithms have been devised to retrieve a signal (up to translation)  from its (possibly noisy) bispectrum~\citet{Giannakis1992}. We therefore find this representation useful in determining the rotationally invariant distances between any pair of images. Bispectrum and triple-correlation function have been considered before for generating translational or rotational invariant features for cryo-EM images~\citet{vanHeel1990a, Joyeux2002, Marabini1996}. However, because the number of such features is extremely large, it was regarded impractical for computations. We reduce the number of bispectrum-features in two steps. We first perform principal component analysis (PCA) for all the images and their in-plane rotations efficiently to produce a steerable basis, where the eigen-images are separable to angular Fourier modes and radial functions~\citet{Zhao2013}. The projection images are expanded and compressed in the leading $M$ steerable eigen-images. Different triplets of these expansion coefficients are multiplied together to produce the invariant image representation. The resulting invariant representation is still high-dimensional, consisting of $O(M^3/k_{max})$ features, where $k_{max}$ is the maximum angular frequency. Marabini and Carazo~\citet{Marabini1996} suggested projecting the bispectrum onto a lower dimensional subspace as a pattern classification method. However, their method consists of using a predetermined subset of bispectrum coefficients and does not preserve the information content well enough to discriminate images of many different views. Instead, in the second step, we reduce the dimensionality of the invariant feature vectors by PCA. We use a randomized algorithm for low rank matrix approximation~\citet{Rokhlin2009, HMT2011, Halko2011} to efficiently compute the principal components, overcoming the difficulties imposed by the large number of images and the high dimensionality of the input feature vectors. The top principal components provide the reduced invariant image representation. We then efficiently compute the rotationally invariant distance between images as the Euclidean distance between their reduced invariant representations without performing any in-plane alignment. A predetermined number of nearest neighbors for each image are identified as those images with the smallest invariant distances. For a large number of input images, a randomized nearest neighbor algorithm~\citet{Osipov2010} can avoid computing the distances between all pairs of images and effectively find the nearest neighbors in time nearly linear with the number of images. Either ordinary or randomized nearest neighbor search with reduced invariant image representation gives the initial classification result. The rotational alignment angles are then computed only for nearest neighbor pairs. With the techniques we propose here, a substantial gain in computation time is obtained by reversing the order of alignment and classification.

\begin{figure}
\begin{center}
\includegraphics[width=0.8\textwidth]{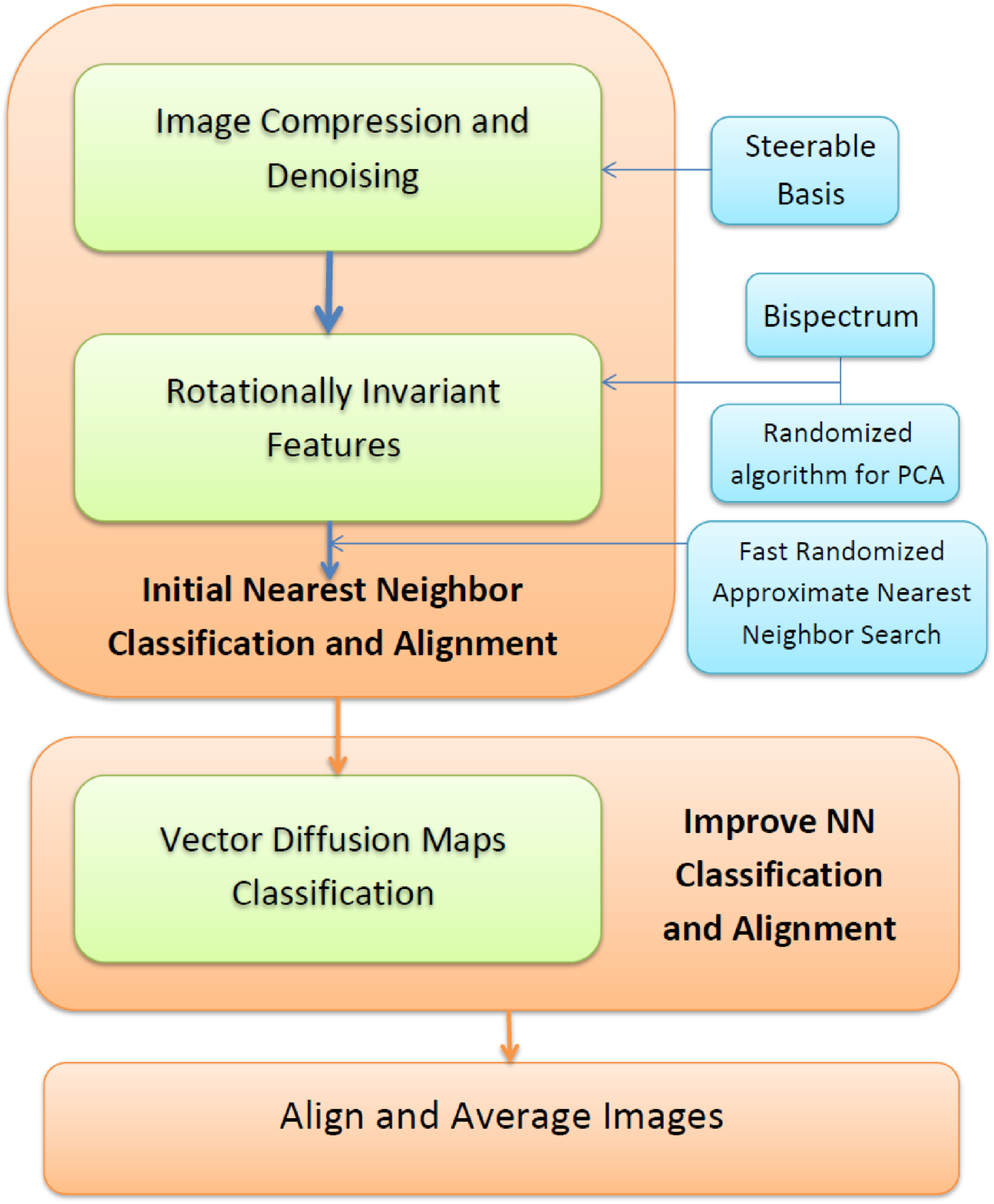}%
\end{center}
\caption{Schematic diagram of our class averaging procedure for single particle reconstruction.}\label{fig:flow_chart}
\end{figure}

The initial nearest neighbors classification can be improved by a clustering algorithm, such as K-means, that takes into account all pairwise distances between images within the neighborhood. But it is usually very difficult to get good clustering for a large number of clusters and the cluster size varies considerably. Nearest neighbor classification is a natural algorithmic framework for averaging an image with a predetermined number of similar images. The initial classification can be further improved by taking into account the consistency of in-plane rotations along multiple paths that connect neighboring images through their common neighbors. This classification method is called Vector Diffusion Maps (VDM)~\citet{Singer2011b, Wu2011}.

This paper is organized as follows. In Section~\ref{sec:Motivation}, we put forward two problems with the reference-free alignment and rotationally invariant K-means clustering. In Section \ref{sec:method}, we present our algorithms for generating rotationally invariant image representations for the purpose of viewing angle classification. Also in that section, we show how to improve the initial nearest neighbor classification and rotational alignment using VDM. The nearest neighbor pairs and their relative alignment are used to generate class means. In Section~\ref{sec:exp}, we detail the results of numerical experiments for simulated projection images of the 70S ribosome with the purpose of benchmarking the efficiency and accuracy of the algorithm. Our algorithm is shown to be more accurate than other existing 2D class averaging procedures and it is also faster. We conclude that section by detailing the results of our class averaging method for three experimental data sets of the 70S ribosome, 50S ribosomal subunit, and IP$_3$R1. Our class averaging method is available in the SPR toolbox ASPIRE\footnote{\url{http://spr.math.princeton.edu/}.}. The toolbox includes three main functions written in MATLAB ``Initial$\_$classification.m'', ``VDM.m'', and ``align\_main.m'' that correspond to the three major components in the pipeline of our 2D class averaging method (see Figure~\ref{fig:flow_chart}). 

\section{Motivation}
\label{sec:Motivation}
\subsection{No global rotational alignment}
\label{subsec:Global_align}
\begin{figure}[h!]
\begin{center}
\includegraphics[width=0.23\textwidth]{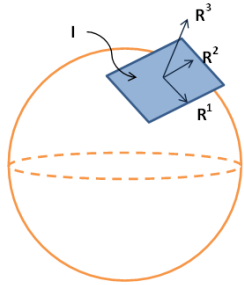}%
\end{center}
\caption{The image $I$ is identified with the tangent plane to the sphere at the viewing direction $R^3$ which is the third column of the rotation matrix $R$.}\label{fig:illustration}
\end{figure}

To each projection image $I$ there corresponds a $3 \times 3$ unknown rotation matrix $R$  ($RR^T = R^TR = I_{3 \times 3}$ and $\det R = 1$), describing its orientation
\[ R = \left( \begin{array}{ccc}
| & | & | \\
R^1 & R^2  & R^3 \\
| & | & | \end{array} \right).\]
The projection image can be viewed as a tangent plane to the two dimensional unit sphere $S^2$ at the viewing direction $v = v(R) = R^3 $. The first two columns of $R$, namely, $R^1$ and $R^2$, are vectors in $\mathbb{R}^3$ that form an orthonormal basis for the tangent plane and are identified with the coordinate axes of the image (see Figure~\ref{fig:illustration}). Together with the imaging direction $v$ they make an orthonormal basis of $\mathbb{R}^3$. An in-plane rotation of the projection image can thus be viewed as changing the basis vectors $R^1$ and $R^2$ while keeping $v$ fixed.

The similarity of images can be measured by the Euclidean distance between the images when they are optimally aligned with respect to in-plane rotations (assuming the images are centered):
\begin{equation}
\label{eq:rotinv_d}
d_{i j} = \min_{\alpha \in [0, 2\pi)} \| I_i - R(\alpha)I_j \|, \quad i, j=1, ..., n,
\end{equation}
where $R(\alpha)$ stands for rotating image $I_j$ counter-clockwise by $\alpha$. The optimal alignment angle is
\begin{equation}
\label{eq:theta}
\alpha_{ij} = \argmin_{\alpha \in [0, 2\pi)} \| I_i  - R(\alpha)I_j \|, \quad i, j = 1, ..., n.
\end{equation}

When two images $I_i$ and $I_j$ are of the same viewing angle ($v_i= v_j$), the matrix $R_i^{-1}R_j$ is of the form
\[ R_i^{-1}R_j = \left( \begin{array}{ccc}
\cos \alpha_{ij} & -\sin \alpha_{ij} & 0 \\
\sin \alpha_{ij} & \cos \alpha_{ij}  & 0 \\
0 & 0 & 1 \end{array} \right),\]
given by $\cos(\alpha_{ij}) = (R_i^{-1}R_j )_{11}$ and $\sin(\alpha_{ij}) = (R_i^{-1}R_j )_{21}$. In practice, however, we cannot expect two projection images to have exactly the same viewing angle.

For clean images, it is expected that a small discrepancy between $v_i$ and $v_j$ would imply that $\alpha_{ij}$, obtained from optimal rotational alignment, approximates the angle $\tilde{\alpha}_{ij}$ given by
\begin{equation}
\label{eq:tilde_theta}
\tilde{\alpha}_{ij} = \argmin_{\alpha \in [0, 2\pi)} \| \rho (\alpha) - R_i^{-1}R_j \|^2_F,
\end{equation}
where
\[ \rho (\alpha) = \left( \begin{array}{ccc}
\cos \alpha & -\sin \alpha & 0 \\
\sin \alpha & \cos \alpha  & 0 \\
0 & 0 & 1 \end{array} \right),\]
and $\|A\|_F^2 = \operatorname{Tr}(AA^T)$ for any real valued $ m \times n$ matrix $A$ (i.e., it is the squared Frobenius norm).
It can be verified that $\tilde{\alpha}_{ij}$ satisfies~\citet{Singer2011b}
\begin{align}
\cos(\tilde{\alpha}_{ij}) &= \frac{(R_i^{-1} R_j)_{11}+(R_i^{-1}R_j)_{22}}{\sqrt{[(R_i^{-1}R_j)_{11}+(R_i^{-1}R_j)_{22}]^2+[(R_i^{-1}R_j)_{21}-(R_i^{-1}R_j)_{12}]^2}}, \label{eq:tilde_theta1}\\
\sin(\tilde{\alpha}_{ij}) &= \frac{(R_i^{-1} R_j)_{21}-(R_i^{-1}R_j)_{12}}{\sqrt{[(R_i^{-1}R_j)_{11}+(R_i^{-1}R_j)_{22}]^2+[(R_i^{-1}R_j)_{21}-(R_i^{-1}R_j)_{12}]^2}}.
\label{eq:tilde_theta2}
\end{align}
During our simulations, the true relative in-plane rotation is defined through equations \eqref{eq:tilde_theta1} and \eqref{eq:tilde_theta2}.

Penczek \textit{et al.}~\citet{Penczek1992} introduced reference-free alignment that first globally aligns all the images and then the rotationally invariant distance is the Euclidean distance between the pre-aligned images. What we are about to elucidate is that such global alignment does not exist when there is a great variety of viewing angles. In such cases, the estimation of the in-plane rotations between images from similar views by RFA is not accurate.

\begin{figure}[h!]
\begin{center}
\subfloat[whole sphere]{
\includegraphics[width=0.3\textwidth]{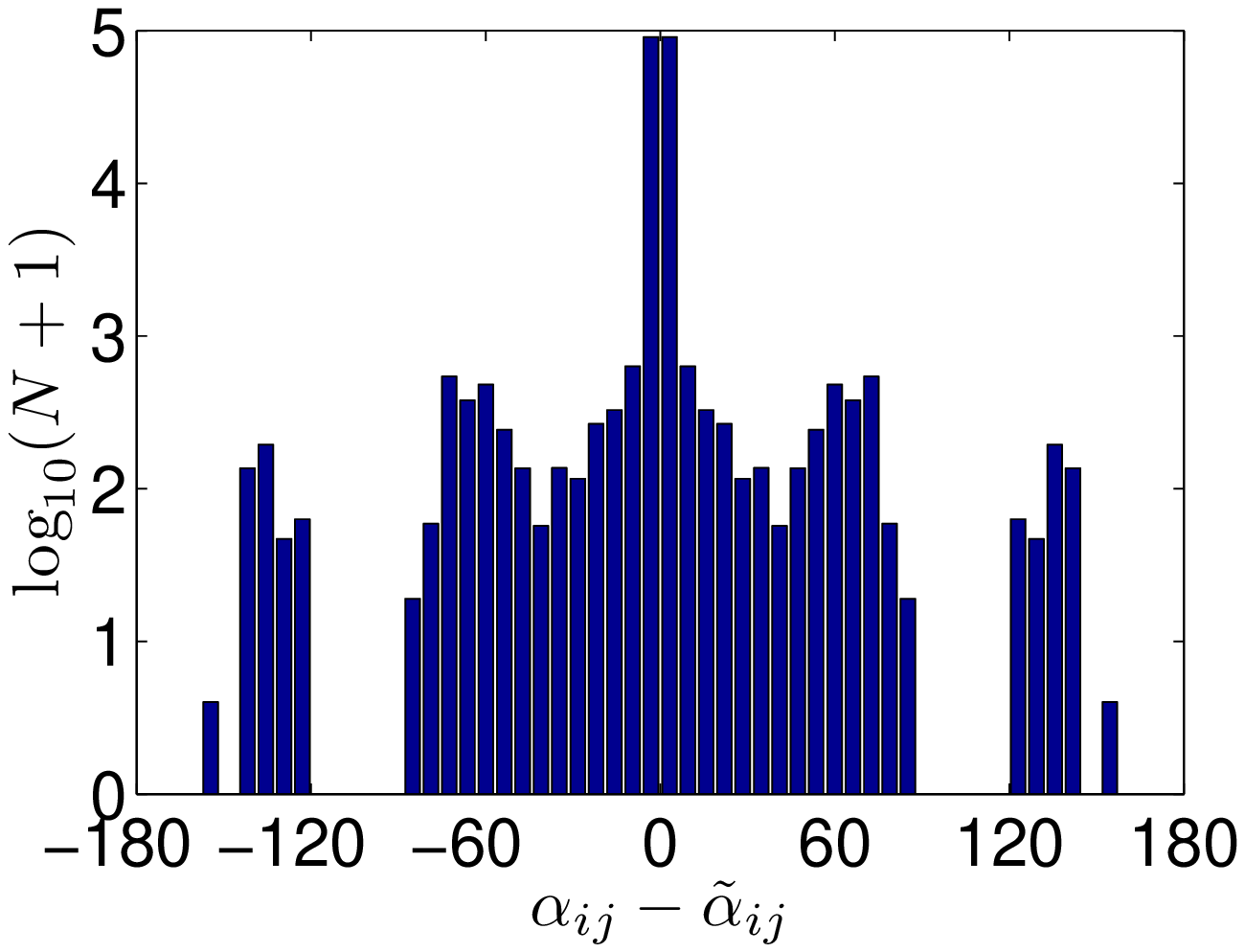}%
\label{fig:whole_sphere}
}
\subfloat[$60\,^{\circ}$]{
\includegraphics[width=0.3\textwidth]{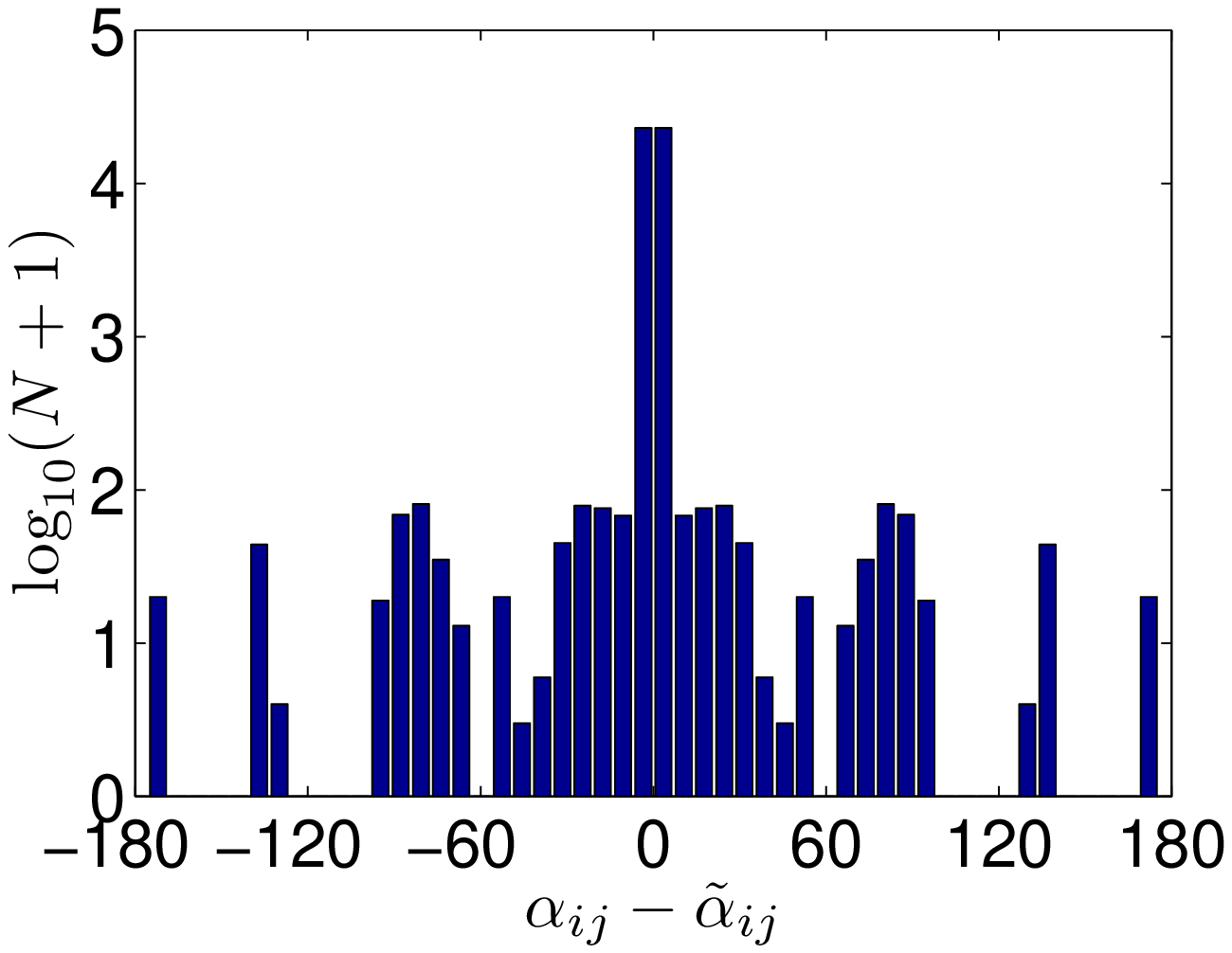}%
\label{fig:60}
}
\subfloat[ $20\,^{\circ}$ ]{
\includegraphics[width=0.3\textwidth]{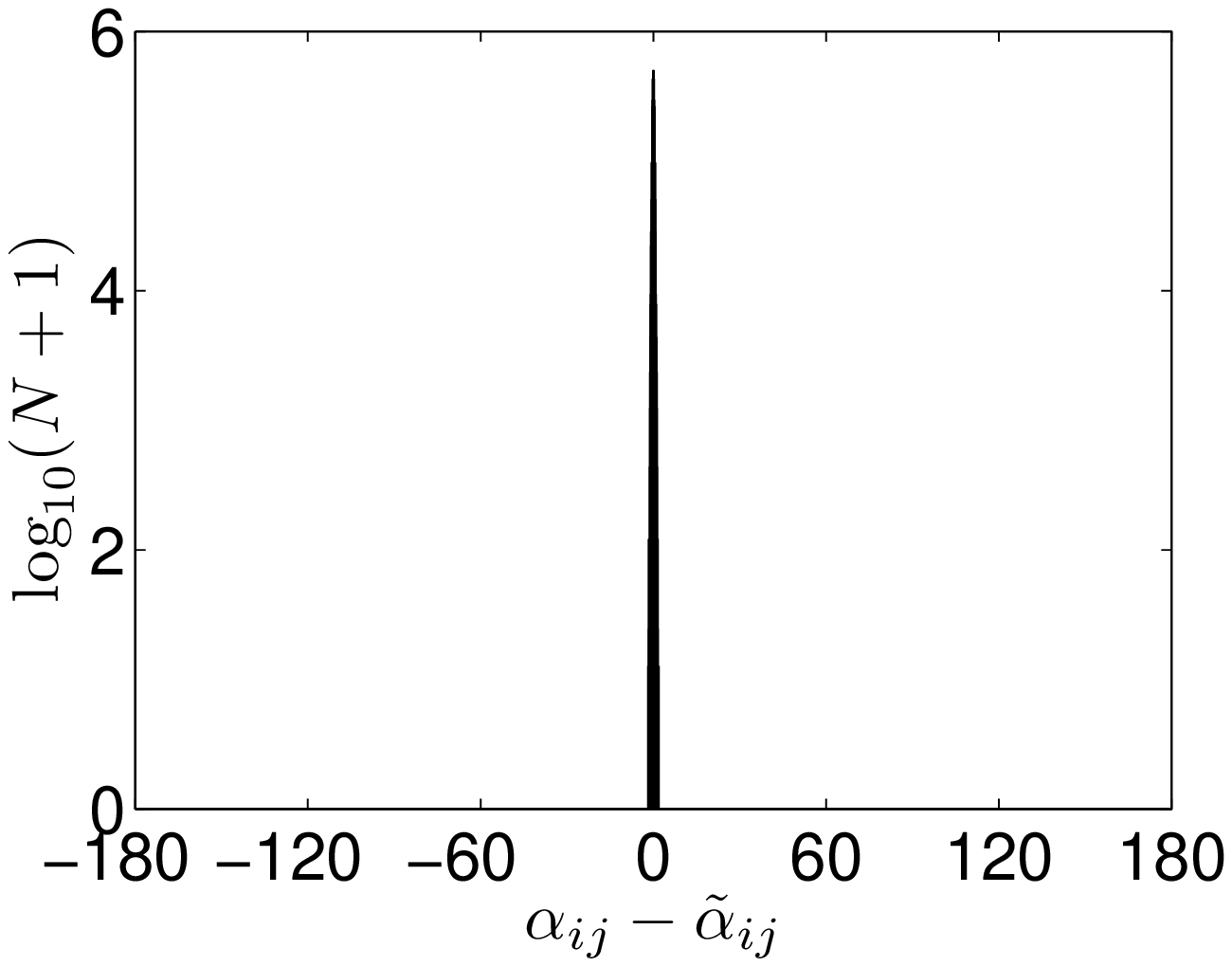}%
\label{fig:20}
}
\end{center}
\caption{Error in degrees of in-plane rotational alignment between images with similar viewing angles that are less than five degrees apart for simulated clean  projection images of the 70S ribosome, with viewing angles belonging to spherical caps of various opening angles (whole sphere, 60 degrees, 20 degrees). The $y$ axis is in $\log$ scale, because the number of outliers is small. The fraction of pairs for which the error is larger than 2 degrees is $p_a=0.13$, $p_b=0.09$, and $p_c=0$.}
\label{fig:SPIDER_error_clean}
\end{figure}

We used a data set composed of clean images corresponding to many different views in order to numerically test the performance of RFA algorithm~\citet{Penczek1992} for viewing angle classification and for rotational alignment of in-class images. Specifically, $10^4$ centered clean projection images were simulated from the 3D model of \textit{Escherichia coli} 70S ribosome with viewing directions that are sampled from the uniform distribution over the sphere. We used SPIDER AP RA program to run RFA on different subsets of the simulated data to test the rotational alignment results. Since we know the underlying rotations, we can compute $\tilde{\alpha}_{ij}$ for pairs of images that satisfy $ \langle v_i, v_j \rangle \geq \cos (5^\circ)$, that is, for viewing angles that are less than $5^{\circ}$ apart. This list of true in-plane rotational angles are compared with the estimation from the reference free alignment. Firstly we ran RFA on the whole data set whose viewing directions are uniformly distributed over the sphere. The algorithm produces large errors when all views are included (see Figure~\ref{fig:whole_sphere}). As we decrease the size of the spherical cap to $80^\circ$ , $60^\circ$ and $40^\circ$, the errors in in-plane rotational alignment become smaller (see Figure~\ref{fig:SPIDER_error_clean}).

The (perhaps surprising) failure of RFA to globally align all images is a consequence of a mathematical theorem called the hairy ball theorem~\citet{Milnor1978}. The theorem says that a continuous tangent vector field to the two dimensional sphere $S^2$ must vanish at some point on the sphere. In other words, if $f$ is a continuous function that assigns a vector in $\mathbb{R}^3$ to every point $v$ on the sphere such that $f(v)$ is tangent to the sphere at $v$, then there is at least one $v \in S^2 $ such that $f(v)=\mathbf{0}$. The theorem attests to the fact that it is impossible to comb a hairy (spherical) cat without creating a cowlick. The hairy ball theorem implies that any attempt to find a non-vanishing continuous tangent vector field to the sphere would ultimately fail. A successful global rotational alignment of all projection images means that we can choose orthogonal bases to all tangent planes such that the basis vectors vary smoothly from one tangent plane to the other. However, this is a contradiction to the hairy ball theorem.

This implies that any classification algorithm that first attempts to globally align the images, such as K-means clustering after RFA, would ultimately fail whenever there are many different views that cover the sphere. We refer the reader to Appendix B of~\citet{Singer2011b} for a discussion about the relevance of the hairy ball theorem in the discrete case of a finite number of images. For images that lie in a spherical cap, the error produced by global alignment is due to the curvature of the sphere.

Since we cannot align images from different views all at once, the distance computed between images after global alignment is not a truly rotationally invariant distance. In Section \ref{sec:method}, we introduce a new rotationally invariant image representation $\tilde{\textbf{b}}$ and replace the rotationally invariant distance \eqref{eq:rotinv_d} by
\begin{equation}
\label{eq:new_rotinv}
d_{ij} = \| \tilde{\textbf{b}}_i- \tilde{\textbf{b}}_j\|.
\end{equation}
The new rotationally invariant feature vector $\tilde{\textbf{b}}$ needs to be lower dimensional (so that \eqref{eq:new_rotinv} can be computed efficiently), and to retain the information in the image (so that \eqref{eq:new_rotinv} is meaningful). Using the rotationally invariant feature vectors, we are able to find images with similar views without performing rotational alignment.

\subsection{Classification instead of clustering}
\label{subsec:Classification_Clustering}
\begin{figure}[h!]
\begin{center}
\includegraphics[width=0.4\textwidth]{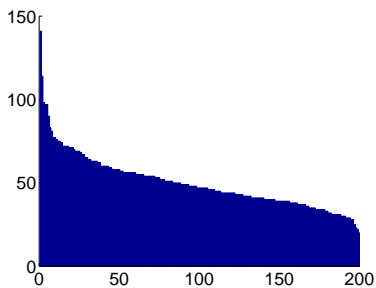}%
\end{center}
\caption{Number of particles in each cluster of the 200 clusters found by K-means clustering algorithm implemented in SPIDER. The data set has $10^4$ clean centered images, whose viewing angles are uniformly distributed over the sphere.} \label{fig:SPIDER_clean_cluster}
\end{figure}

Traditionally, the class averaging problem was considered as a clustering problem, in which a large data set of $n$ images, $I_1, ... ,  I_n$ with unknown corresponding rotation matrices $R_1,...,R_n$, is grouped into clusters, with the goal that images within a single cluster have similar viewing angles. In practice, however, the size of the cluster varies considerably from cluster to cluster (see Figure \ref{fig:SPIDER_clean_cluster}, where we tried to cluster $10^4$ clean 70S ribosome projection images, whose viewing angles are uniformly distributed over the sphere), and therefore the resulting class averages will have different signal to noise ratio and resolution.

Instead of K-means clustering and generating cluster means, we propose another classification method for generating class averages. For each image, we search for a fixed number ($\kappa$) of nearest neighbors. Each image is averaged with its aligned nearest neighbors to boost the signal to noise ratio. In this way, the resulting number of class averages is the same as the number of the original images and all class averages have the same SNR. It also prevents the situation that clustering reduces the full coverage of the viewing directions.

\section{Methods}
\label{sec:method}
\subsection{Fourier-Bessel Steerable PCA}
\label{subsec:FBsPCA}
We use Fourier-Bessel steerable PCA~\citet{Zhao2013} to generate a data adaptive basis for compressing and de-noising images. Since the rotated copies of the projection images are equally likely to appear in the data set, it is meaningful to perform PCA on the data set with all their rotated copies. However it is challenging to compute the steerable PCA efficiently and accurately, because the images are sampled on a Cartesian grid, while steering operations often require a polar grid. As the transformation from Cartesian to polar is not unitary, the eigenimages corresponding to images mapped to polar grid are not equivalent to transforming the original eigenimages from Cartesian to polar. In~\citet{Zhao2013}, we developed an accurate and efficient algorithm for steerable PCA. Its computational complexity is lower than that of traditional PCA (or MSA in cryo-EM 2D image processing). Since we incorporate more information from the data set, we can get better estimation of eigen-images that correspond to the clean projection images than the traditional PCA. 

The steerable eigen-images have special separation of variables form, 
\begin{equation}
\label{eq:steerable_basis}
u^{k, q}(r, \theta) = f^{k, q} (r) e^{\iota k \theta},
\end{equation}
where $k$ and $q$ in basis image $u^{k, q}$ are indices for angular frequency and radial frequency, respectively. $f^{k, q}$ can be computed from the Fourier-Bessel steerable PCA~\citet{Zhao2013}, which provides an optimal basis in the least-squares sense. Images are expanded on this steerable basis, $I(r, \theta) = \sum_{k, q}a_{k, q}u^{kq}(r, \theta)$, with expansion coefficients $a_{k, q}$. It is easy to ``steer'' the images.  When image $I$ is rotated counter-clockwise by angle $\alpha$, the expansion coefficients of $I(r, \theta-\alpha)$ are given by $a^\alpha_{k, q} =  a_{k, q} e^{-\iota k \alpha}$, because
\begin{align}
\label{eq:Ii}
I(r, \theta-\alpha) &= \sum_{k, q} a_{k, q} u^{kq}(r, \theta-\alpha) \nonumber \\
&= \sum_{k, q} a_{k,q} e^{-\imath k \alpha} u^{kq} (r, \theta).
\end{align}
The steerability of the basis allows us to define rotationally invariant features that are introduced in Section~\ref{subsec:bispec}. 

\subsection{Bispectrum-like Rotationally Invariant Image Representation}
\label{subsec:bispec}
Prior to introducing the rotationally invariant image representation, we quickly review here the bispectrum for 1D signals. Suppose we have a 1D periodic discrete signal $f(x), x=1, ..., L $. The discrete Fourier transform of $f$ is defined as
\begin{equation}
\label{eq:Fourier Transform}
\hat{f} (k) = \sum_{x=1}^{L} f(x) e^{-i \frac{2\pi}{L}k x}.
\end{equation}
The power spectrum $|\hat{f} |^2$ is the Fourier transform of the autocorrelation function
\begin{equation}
ACF(x)=\sum_{y=1}^{L} \overline{f(y)} f(y+x) .
\end{equation}
Both the power spectrum and the auto-correlation function are shift-invariant. However, the ACF loses the phase information in $\hat{f}$ and maintains only its amplitude. The idea behind bispectral invariants is to move from the autocorrelation function to the triple-correlation function
\begin{equation} \label{eq:triple_correlation}
T(x_1, x_2) = \sum_{y=1}^{L} \overline{f(y)}f(y+x_1) f(y+x_2).
\end{equation}
Again by the convolution theorem, Fourier transform of the triple-correlation function is
\begin{equation}
\label{eq:bispec}
b(k_1, k_2) = \hat{f}(k_1) \hat{f}(k_2) \overline{\hat{f}(k_1+k_2)},
\end{equation}
and is called the bispectrum of $f$. Under shift by $z$, the Fourier transform of $f^z = f (x-z)$ becomes
\begin{align}
\label{eq:FFT_shift}
\hat{f}^z(k) &= \sum_{x=1}^L f(x-z)\, e^{-i \frac{2\pi}{L}k x} =  e^{- i \frac{2\pi}{L}k z} \sum_{x'=1}^L f (x')\, e^{-i \frac{2\pi}{L}k x'} = e^{- i \frac{2\pi}{L}k z} \hat{f}(k).
\end{align}
Therefore, under translation by $z$, the bispectrum becomes
\begin{equation}
\label{eq:bshift}
b^z(k_1, k_2) = e^{-i 2 \pi z k_1 /L} \hat{f}(k_1) e^{- i 2 \pi z k_2/L } \hat{f}(k_2) e^{i 2 \pi z (k_1+k_2)/L} \overline{\hat{f}(k_1+k_2)} = b(k_1, k_2),
\end{equation}
which shows that the bispectrum is shift-invariant. Unlike the power spectrum, the bispectrum does not lose the phase information and under mild conditions, the original signal can be reconstructed from its bispectrum (up to translation). The bispectrum is widely used in signal processing as a lossless shift-invariant representation, and various algorithms have been devised to reconstruct $f$ from $b$ \citet{Giannakis1992}. Because of the symmetry properties of bispectrum coefficients, the knowledge of the bispectrum in the triangular region $k_1\geq 0$, $k_2 \leq k_1$, $k_1+k_2 \leq k_{max}$ is sufficient for a complete description of the bispectrum.

For $1$D periodic signals of length $L$, there are $O(L^2)$ bispectrum coefficients. Therefore, the bispectrum is of very high dimensionality. The possibility of using the bispectrum as shift or rotational invariant image representation for classification of cryo-EM images has been previously mentioned in~\citet{vanHeel1990a,Marabini1996}. Due to its high dimensionality, the full bispectrum has never been used for analyzing large cryo-EM data sets to generate class averages.

The bispectrum of 1D periodic signals for shift invariant features can be extended to generate rotationally invariant features for 2D images. We use Fourier-Bessel steerable PCA basis~\citet{Zhao2013} described in Section~\ref{subsec:FBsPCA} to expand images. Rotating the image is equivalent to phase shifting its expansion coefficients, which is similar to phase shifting the Fourier coefficients in \eqref{eq:FFT_shift}.

Typically, most of the energy of the clean images is concentrated in a relatively small number $M$ (a typical value of $M$ is around $100$ for noisy 2D images) of principal components with low angular frequencies ($-k_{max} \leq k \leq k_{max}$), whereas the additive white Gaussian noise spreads over all components with low angular frequencies. Representing the images using only the leading $M$ components can compress and denoise the images. Therefore, we use the truncated expansion coefficients with $M$ terms instead of the total number of pixels.

We define the bispectrum for the steerable basis expansion coefficients as
\begin{equation} \label{eq:bispec2D}
b_{k_1, k_2, q_1, q_2, q_3} = a_{k_1, q_1} a_{k_2, q_2}\overline{a_{k_1+k_2, q_3}},
\end{equation}
where $k_1$ and $k_2$ are the angular indices and $q_1$, $q_2$ and $q_3$ are the radial indices.

A modification to the bispectrum is needed when treating noisy signals. Suppose the observed signal $y$ is the true signal $x$ contaminated with additive white Gausfdsian noise $\mathfrak{n} \sim \mathcal{N}(0,\sigma^2 I)$:
\begin{equation}
y = x + \mathfrak{n}.
\end{equation}
Then the expansion coefficients are given by
\begin{equation}
a^y_{k, q} = a^x_{k, q}+a^\mathfrak{n}_{k, q},
\end{equation}
with $a^\mathfrak{n}_{k, q}$ satisfying $\mathbb{E}a^\mathfrak{n}_{k, q}=0$ and $\mathbb{E} [a^\mathfrak{n}_{k_1, q_1} \overline{a^\mathfrak{n}_{k_2, q_2}}] = \sigma^2 \delta_{k_1 k_2}\delta_{q_1 q_2}$.
Then the expectation of the bispectrum of $y$,
\begin{align}
\mathbb{E}b^y_{k_1, k_2, q_1, q_2, q_3} &= \mathbb{E} [(a^x_{k_1, q_1}+a^\mathfrak{n}_{k_1, q_1})(a^x_{k_2, q_2}+a^\mathfrak{n}_{k_2, q_2})(\overline{a^x_{k_1+k_2, q_3}+a^\mathfrak{n}_{k_1+k_2, q_3}})] \nonumber \\
& = a^x_{k_1, q_1} a^x_{k_2, q_2} \overline{a^x_{k_1+k_2, q_3}}+ \mathbb{E} \left[a^\mathfrak{n}_{k_1, q_1} a^\mathfrak{n}_{k_2, q_2} \overline{a^\mathfrak{n}_{k_1+k_2, q_3}} \right] \nonumber  \\
& \quad + a^x_{k_2, q_2}\overline{a^x_{k_1+k_2, q_3}}\mathbb{E}\left[a^\mathfrak{n}_{k_1, q_1}\right] + a^x_{k_1, q_1}\overline{a^x_{k_1+k_2, q_3}} \mathbb{E}\left[a^\mathfrak{n}_{k_2, q_2}\right] \nonumber \\
& \quad + a^x_{k_1, q_1}a^x_{k_2, q_2}\mathbb{E}\left[\overline{a^\mathfrak{n}_{k_1+k_2, q_3}}\right] \nonumber  + a^x_{k_1, q_1} \mathbb{E}\left[ a^\mathfrak{n}_{k_2, q_2}\overline{a^\mathfrak{n}_{k_1+k_2, q_3}}\right] \nonumber \\
& \quad + a^x_{k_2, q_2}\mathbb{E}\left[ a^\mathfrak{n}_{k_1, q_1}\overline{a^\mathfrak{n}_{k_1+k_2, q_3}}\right]+\overline{a^x_{k_1+k_2, q_3}} \mathbb{E} \left[a^\mathfrak{n}_{k_1, q_1}a^\mathfrak{n}_{k_2, q_2} \right].
\end{align}
Hence,
\begin{equation}
\mathbb{E}b^y_{k_1, k_2, q_1, q_2, q_3} = b^x_{k_1, k_2, q_1, q_2, q_3} + \sigma^2 (\delta_{q_2, q_3}a^x_{0, q_1}+\delta_{q_1, q_3}a^x_{0, q_2}+\delta_{q_1, q_2}a^x_{0, q_3}).
\end{equation}
Therefore, if $a^x_{0, q}=0$ for all $q$, then the bispectrum is unbiased, i.e., $\mathbb{E} \textbf{b}^y = \textbf{b}^x$. As a result, removing the zero-frequency part of the bispectrum makes it less sensitive to contamination by additive white Gaussian noise. The zero-frequency coefficients are rotational invariant and can be added as separate invariant features. 

Van Heel \textit{et al.}~\citet{vanHeel1990a, vanHeel1992} have previously noted that the  ACF overweighs the already strong frequency components in the image due to the squaring of the Fourier components and therefore they defined a self correlation function (SCF) which under-emphasizes all amplitudes by replacing them by their square roots.  The SCF was shown to perform better than the ACF. A similar situation occurs for the bispectrum, due to the multiplication of three frequency components. We therefore modify the expansion coefficients prior to computing the bispectrum such that the amplitude is the cubic root of the original:
\begin{align}
\tilde{a}^i_{k, q} = \frac{a^i_{k, q}}{|a^i_{k, q}|^{2/3}}.
\label{eq:CubicRoot}
\end{align}
Notice that the phase information of the bispectrum is unaltered, as only the amplitudes are modified. It is natural to take the cubic root since in this way the bispectrum scales linearly with the intensity of the image (that is, multiplying an image $I$ by the constant $c$ results in multiplication of $b$ by $c$, instead of $c^3$ for $b$).

The rotationally invariant image representation derived in \eqref{eq:bispec2D} is of very high dimensionality. Suppose that the truncated expansion coefficients have $M$ components and that the corresponding maximum angular frequency is $k_{max}$, then the resulting invariant feature vector is of length $O\left(\frac{M^3}{k_{max}}\right)$. Computing the inner product of vectors of length $10^4-10^5$ can be quite expensive. It is therefore required to reduce the dimensionality of the invariant feature vectors. While this reduction can be achieved by PCA, the typically large number of images and the high dimensionality of the feature vectors make the computational cost of classical PCA quite demanding. Instead, we use the recently proposed randomized algorithm for low rank matrix approximation~\citet{Rokhlin2009, HMT2011, Halko2011}. We denote by $M'$ the reduced dimension, that is, the number of principal components chosen in this step.

We define the rotationally invariant affinity between image $I_i$ and image $I_j$ as the normalized cross-correlation $C_{ij}$ between their corresponding low dimensional feature vectors of length $M'$, where $M'$ is about $200$ in application. 

A fixed number of nearest neighbors with the largest normalized cross-correlation $C_{ij}$ with image $i$ are determined, with computational complexity $O(n^2 M')$. For large data sets, consisting of $10^5$ images or more, the randomized approximate nearest neighbor (RANN) search algorithm~\citet{Osipov2010} is an efficient way for finding the nearest neighbors without computing $C_{ij}$ for all pairs of $i$ and $j$. RANN is an iterative algorithm. It first randomly rotates the data points (in our case, complex valued vectors of length $M'$ ) and subdivides them into smaller boxes by looking at $1$, $2$, $3$, $4 \dots$ coordinates, until each box contains about $\kappa$ points. Then the suspected nearest neighbors are determined locally as those in same boxes. The process is repeated through independent iterations, and the list of suspected neighbors is refined.  In practice only a small number of iterations is needed in order to find the true nearest neighbors with very high probability. The computational complexity for this randomized algorithm is $O(Tn(M'\log M'+\kappa \log \kappa \log n)+n \kappa^2 (M'+\log \kappa))$, where $T$ is the number of iterations and $\kappa$ is the number of nearest neighbors.

After classifying images of similar views, we rotationally align images with their nearest neighbors. The in-plane rotation angle $\alpha_{ij}^*$ for a pair of neighboring images $I_i$ and $I_j$ is determined by aligning their denoised versions. 

\subsection{Vector Diffusion Maps Classification and Rotational Alignment}
\label{subsec:VDM}
\begin{figure}[h!]
\begin{center}
\subfloat[]{
\includegraphics[width=0.4\textwidth]{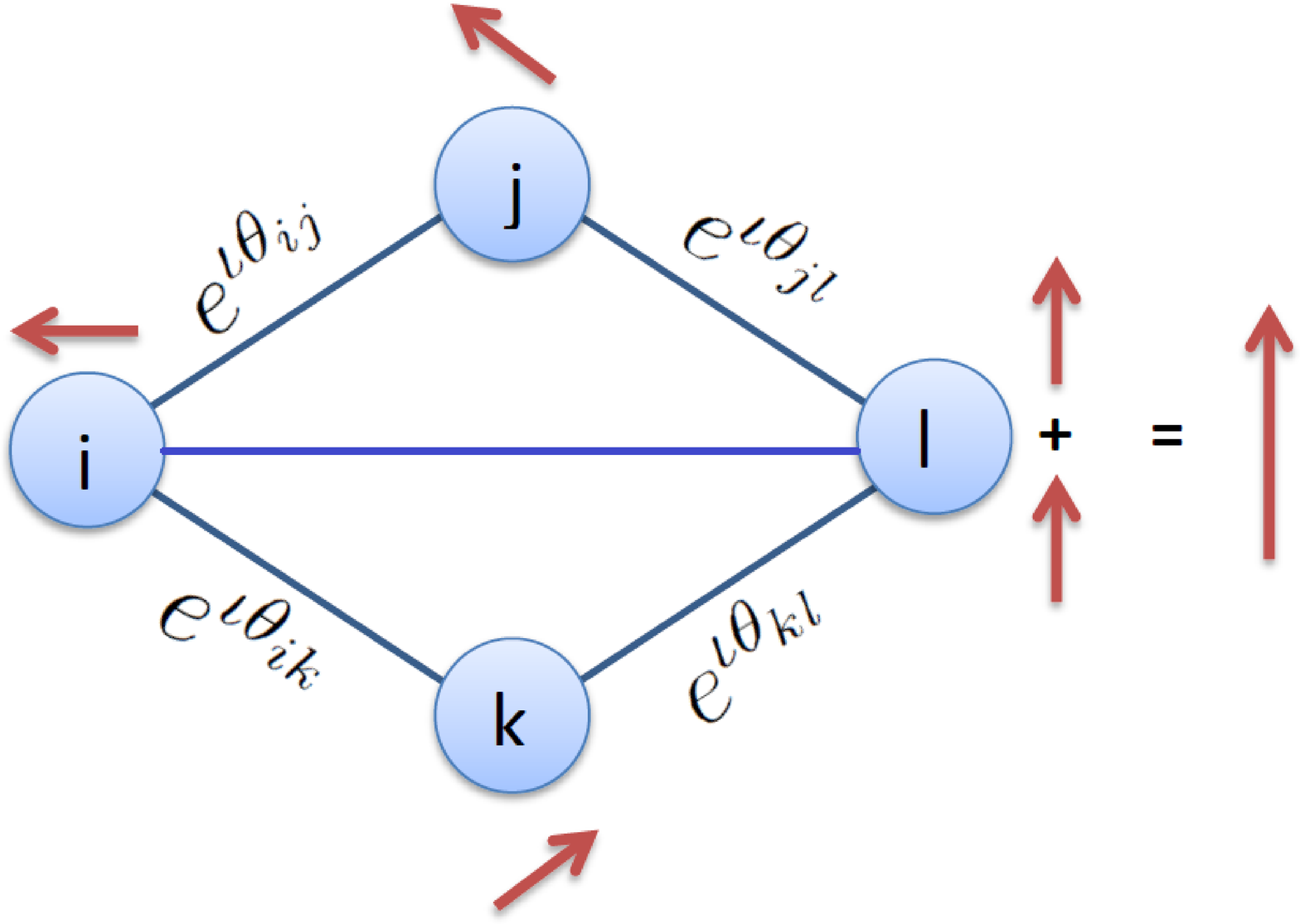}%
\label{fig:ill1}
}\quad
\subfloat[]{
\includegraphics[width=0.4\textwidth]{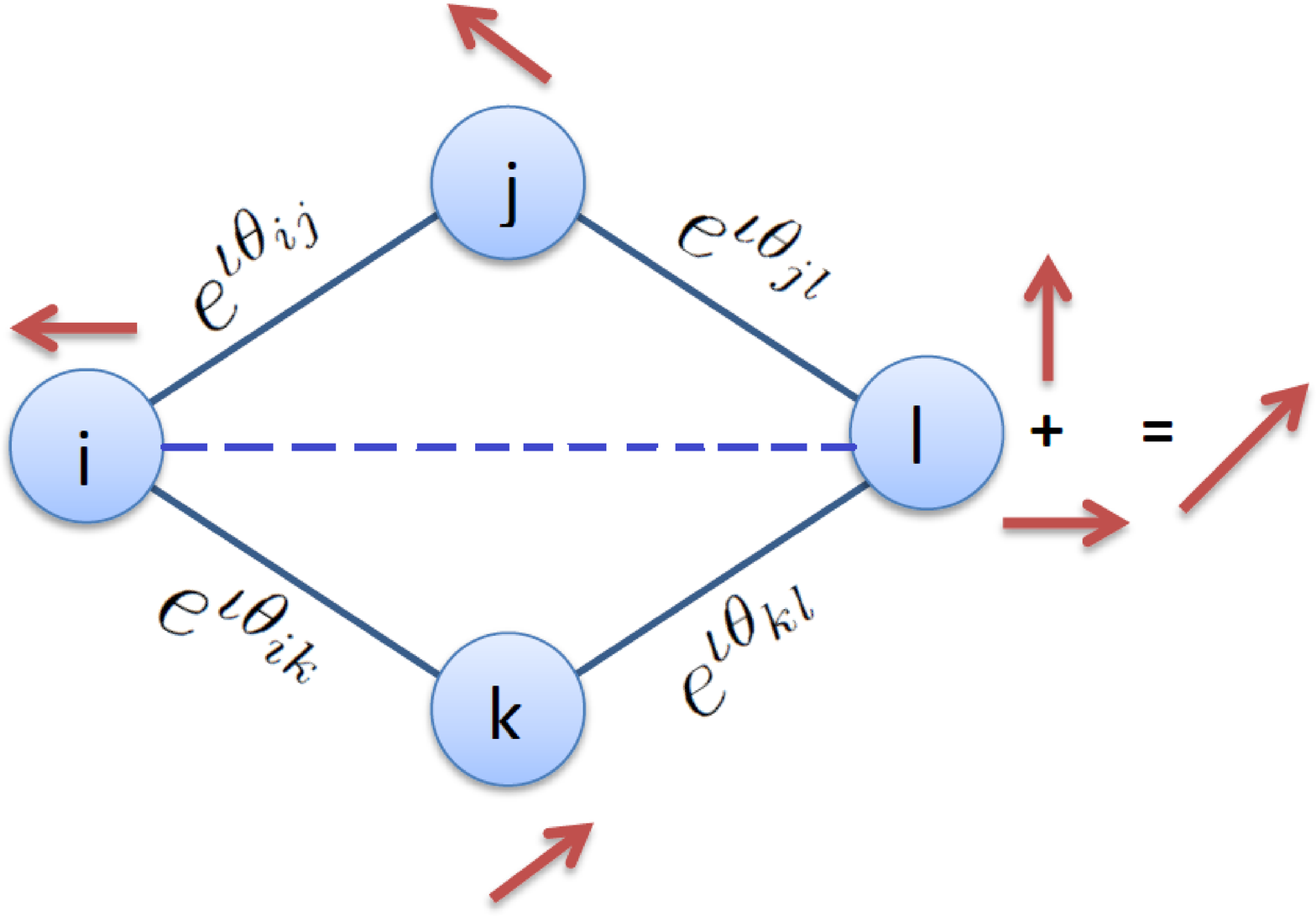}%
\label{fig:ill2}
}
\end{center}
\caption{ Illustration for Vector Diffusion Map (VDM) affinity. Pick an arbitrary planar vector for node $i$ (realized as a complex number $e^{\iota \phi}$). Consider two different paths from $i$ to $l$ of length $2$: $i\rightarrow j\rightarrow l$ and $i\rightarrow k\rightarrow l$. The arrow is rotated according to the edges $i\rightarrow j$ and $i\rightarrow k$, respectively (by multiplying it by the phase factors, $e^{\iota \theta_{ij}}$ and $e^{\iota \theta_{ik}}$, respectively), and then rotated according to the edges $j\rightarrow l$ and $k\rightarrow l$, respectively (by multiplying it by the phase factors, $e^{\iota \theta_{jl}}$ and $e^{\iota \theta_{kl}}$, respectively). Different paths may be consistent as in (a) or inconsistent as in (b). When vectors from different paths are added together the amplitude of the resulting vector can be as large as the number of paths if they are all consistent (a), or much smaller due to inconsistencies (b). Node $i$ and node $l$ have higher affinity in (a) than in (b). } \label{fig:illu_VDM}
\end{figure}

When the SNR is very low, the initial rotationally invariant classification based on just nearest neighbors might still give some outliers. Further improvement can be obtained by taking into account the consistency of pairwise distances and rotational transformations among the images in the neighborhood. This can be achieved by using a classification method called Vector Diffusion Maps (VDM)~\citet{Wu2011, Singer2011b}, which is a generalization of Diffusion Maps, a popular method in manifold learning~\citet{Lafon2006}. This method takes into account the consistency of in-plane rotational transformations (see Figure~\ref{fig:illu_VDM}). The affinity between images $I_i$ and $I_j$ (shown as nodes $i$ and $j$) is defined as the consistency of the transformations summed over all different paths of a fixed length connecting $i$ and $j$. To quantify this, we build a sparse $n\times n$ Hermitian matrix $H$ \eqref{eq:Hmatrix} using the union rule that $i$ and $j$ are neighbors if either $i$ is one of $j$'s $\kappa$ nearest neighbors or $j$ is one of $i$'s $\kappa$ nearest neighbors,
\begin{equation}
H_{ij}=
\begin{cases} e^{\iota \alpha^*_{ij}} & \text{$\{i, j\}\in E$,}\\
0, & \text{$ \{i, j\} \not \in E$,}
\end{cases}
\label{eq:Hmatrix}
\end{equation}
where $E$ denotes the set of neighboring pairs and $\alpha_{ij}^*$ is the optimal in-plane rotation of images $I_i$ and $I_j$.
The fact that $H$ is Hermitian follows from $\alpha_{ij}^* = -\alpha_{ji}^* \mod 2\pi$. Moreover, since only neighboring images contribute non-zero entries in $H$,  it follows that $H$ is a sparse matrix whose storage requires only $O(n\kappa)$ space. Each row of $H$ is divided by the degree of the corresponding image, yielding the matrix $S$ that is given by
\begin{equation}
\label{eq:S}
S = D^{-1} H,
\end{equation}
where $D$ is an $ n \times n$ diagonal matrix with
\begin{equation}
D(i,i)=\operatorname{deg}(i) = \sum_j |H_{ij}|.
\end{equation}
The matrix $S$ \eqref{eq:S} is similar to the Hermitian matrix
\begin{equation}
\tilde{S} = D^{-1/2} H D^{-1/2}
\end{equation}
through $S = D^{-1/2} \tilde{S} D^{1/2}$.
We can define the affinity between $i$ and $j$ as $|\tilde{S}^{2t} (i, j)|^2$, that is, as the squared absolute value of $\tilde{S}^{2t}(i, j)$, which takes into account all paths of length $2t$, where $t$ is a positive integer. In a sense, $|\tilde{S}^{2t} (i, j)|^2$ measures not only the number of paths of length $2t$ connecting $i$ and $j$ but also the amount of agreement between their transformations. That is, for a fixed number of paths, $|\tilde{S}^{2t} (i, j)|^2$ is larger when the path transformations are in agreement, and is smaller when they differ. We define the normalized affinity between $i$ and $j$ as
\begin{equation}
\frac{| \tilde{S}^{2t} (i, j) |^2}{\sqrt{| \tilde{S}^{2t} (i, i) |^2 | \tilde{S}^{2t} (j, j) |^2}}.
\label{eq:normalized_affinity}
\end{equation}
Since $\tilde{S}$ is Hermitian, it has a complete set of eigenvectors $v_1, v_2, ..., v_n$ and real eigenvalues $\lambda_1, \lambda_2, ..., \lambda_n$. We order the eigenvalues in decreasing order of magnitude. The spectral decomposition of $\tilde{S}$ and $\tilde{S}^{2t}$ are given by
\begin{equation}
\tilde{S}(i, j) = \sum_{l=1}^n \lambda_l v_l(i) \overline{v_l(j)}, \quad \textrm{and} \quad \tilde{S}^{2t}(i, j) = \sum_{l=1}^n \lambda_l^{2t} v_l(i) \overline{v_l(j)}. \label{eq:Smatrix}
\end{equation}
It follows that the affinity $|\tilde{S}^{2t} (i, j)|^2$ is an inner product for the finite dimensional Hilbert space $\mathbb{C}^{n^2}$ via the mapping $V_t$ :
\begin{equation}
V_t : i\mapsto (( \lambda_l \lambda_r)^t v_l(i) \overline{v_r(i)})_{l, r = 1}^{n}.
\end{equation}
That is,
\begin{equation}
|\tilde{S}^{2t} (i, j)|^2= \langle V_t(i), V_t(j) \rangle.
\end{equation}
Then the normalized affinity \eqref{eq:normalized_affinity} can be expressed using the mapping $V_t$ as
\begin{equation}
\label{eq:normalized_affinity2}
\frac{| \tilde{S}^{2t} (i, j) |^2}{\sqrt{| \tilde{S}^{2t} (i, i) |^2 | \tilde{S}^{2t} (j, j) |^2}} = \left\langle \frac{V_t(i)}{|V_t(i)|}, \frac{V_t(j)}{|V_t(j)|} \right\rangle.
\end{equation}
The matrix $\tilde{S}^{2t}$ may be too dense to be computed efficiently. Instead, we can approximate the normalized affinity \eqref{eq:normalized_affinity2} by truncating the mapping $V_t$ to its leading $m^2$ coordinates (instead of $n^2$) as
\begin{equation}
\label{eq:approx_map}
V^m_t : i\mapsto (( \lambda_l \lambda_r)^t v_l(i)\overline{v_r(i)})_{l, r = 1}^{m}.
\end{equation}
where $m$ is the largest integer satisfying $\lambda_m^{2t} > \delta$ for some $\delta$ much smaller than $1$. The approximate normalized affinity becomes
\begin{equation}
\label{eq:approx_aff}
\left\langle \frac{V^m_t(i)}{|V^m_t(i)|}, \frac{V^m_t(j)}{|V^m_t(j)|} \right\rangle.
\end{equation}

We use \eqref{eq:normalized_affinity} as the measure of closeness between two images to improve our estimation of the $\kappa$ nearest neighbors for each image. This measure of affinity can be approximated using the eigenvectors of the matrix $\tilde{S}$ as shown in \eqref{eq:approx_map} and \eqref{eq:approx_aff}~\citet{Wu2011}. The algorithm is very efficient in terms of running time and memory requirements, because it is based on the computation of the top eigenvectors of a sparse Hermitian matrix.

\begin{figure}[h!]
\begin{center}
\subfloat[]{
\includegraphics[width=0.3\textwidth]{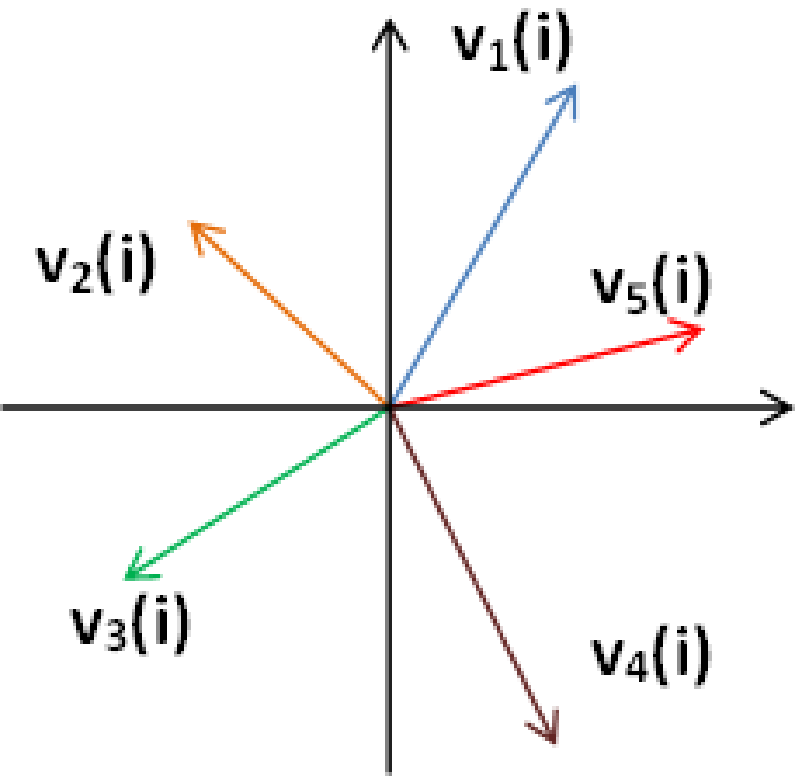}%
\label{fig:rot_align1}
}\quad\quad\quad
\subfloat[]{
\includegraphics[width=0.3\textwidth]{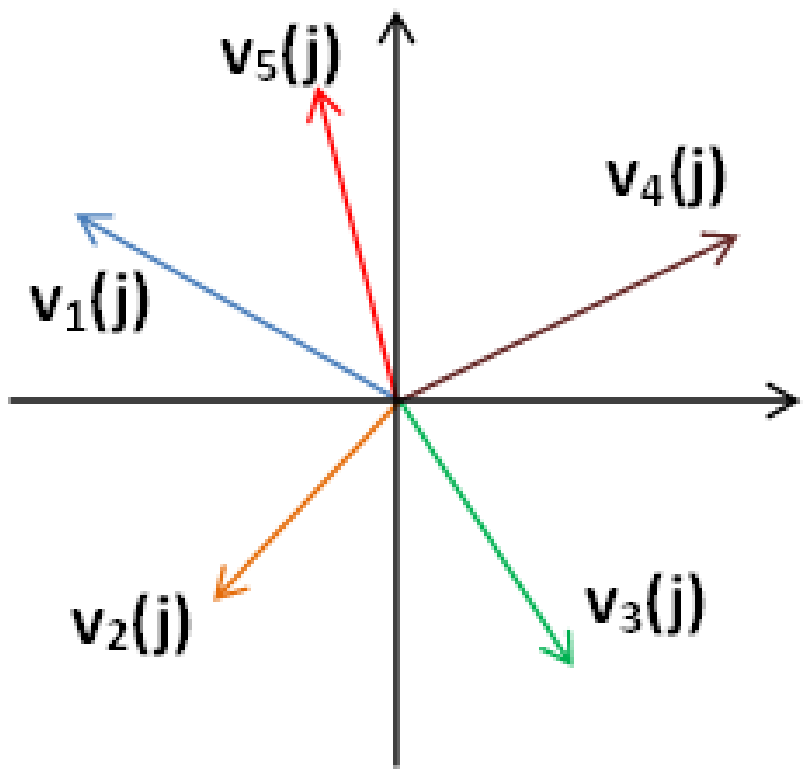}%
\label{fig:rot_align2}
}
\end{center}
\caption{When $i$ and $j$ are of the same viewing angle, the tangent plane at point $i$ coincides with the tangent plane at point $j$ and the eigenvectors satisfy equation \eqref{eq:rot_align_VDM}.} \label{fig:rot_align_VDM}
\end{figure}

The eigenvectors of $\tilde{S}$ encode the information for in-plane rotational alignment between neighboring images. For clean images, if $i$ and $j$ are of the same viewing directions and their in-plane alignment angle is $\alpha_{ij}$, the following holds
\begin{equation}
v_l(i) = e^{\iota \alpha_{ij}} v_l(j), \quad \forall l=1, ..., n.
\label{eq:rot_align_VDM}
\end{equation}
This is illustrated in Figure~\ref{fig:rot_align_VDM}.
When the viewing directions are close (though not identical), then \eqref{eq:rot_align_VDM} holds approximately. The level of approximation deteriorates as the eigenvalues become smaller, because their corresponding eigenvectors are more oscillatory and more sensitive to noise. Therefore, we use $\lambda_l^{2t}$ to give more weight to the leading eigenvectors. We estimate the rotational angle using the top $m$ eigenvectors:
\begin{equation}
\label{eq:VDM_angle1}
\alpha_{ij}^*  = \argmin_{\alpha_{ij}} \sum_{l=1}^m \lambda_l^{2t} | v_l (i) - e^{\iota \alpha_{ij}} v_l (j) |^2, 
\end{equation}
given by
\begin{equation}
e^{\iota \alpha^*_{ij}} = \frac{\sum_{l=1}^m \lambda_l^{2t} v_l(i) \overline{v_l(j)}}{|\sum_{l=1}^m \lambda_l^{2t} v_l(i) \overline{v_l(j)}|}.
\label{eq:VDM_angle2}
\end{equation}
In this way, we improve the estimation of the in-plane rotational alignment between nearest neighbors.

\subsection{Shift Alignment}
\label{subsec:sa}
The experimental particle images are cropped from the micrographs through a particle selection procedure, and therefore they are not centered. Shift alignment is needed for generating class averages. Ideally we would like to center all images before performing rotational alignment and classification. What we are going to elucidate is that it is hard to center all projection images at the class averaging stage.

There are three degrees of freedom in defining the centers of all images. The three degrees of freedom correspond to the definition of the center of the three-dimensional molecule. We can fix the three degrees of freedom by choosing the center of mass of the volume as the origin. Then the center of mass of the clean projection images should also be at the origin. Therefore, for clean images with the same CTF function, we can center the images by finding the center of mass of the projection images. However, this method performs poorly at low SNR and when images are pooled together from different defocus groups. The practical procedure in the field is to shift-align the images iteratively by correlating them with the mean of the data set or with a circular reference image. The estimation error for this procedure is typically of the order of 5 pixels in each direction.

To align images, we have to perform brute force shift search for the rudimentarily shift-aligned images. For image $i$ and image $j$ of the same view, with relative in-plane rotation angle $\alpha^*_{ij}$ and relative shift $(s_{ij, x}, s_{ij, y})$, the following equation holds,
\begin{equation}
 \left(\begin{array}{c} x_i\\ y_i \end{array} \right) -  \left(\begin{array}{cc} \cos\alpha^*_{ij} & -\sin\alpha^*_{ij}\\\sin\alpha^*_{ij}& \cos \alpha^*_{ij}  \end{array} \right)   \left(\begin{array}{c} x_j\\ y_j \end{array} \right)  = \left(\begin{array}{c} s_{ij, x}\\s_{ij, y} \end{array} \right),
\label{eq:shift_1}
\end{equation}
where $(x_i, y_i)$ and $(x_j, y_j)$ are the location of the center of the projection images.
Equation \eqref{eq:shift_1} is exact only when $i$ and $j$ share exactly the same viewing direction. When they are slightly different, this equation is not exact anymore. Therefore, the least squares solution to \eqref{eq:shift_1} does not produce the true global shifts $(x_i, y_i)$. The least squares solution would perform well in aligning neighboring images, but it is not expected to find the shifts between different classes.

The rotationally invariant features described in Section~\ref{subsec:bispec} are not shift invariant. Therefore, we would like the images to be centered. However, as we have shown above, centering the images at the stage of class averaging is hard to achieve. As a result, in practice we use low pass filtering to make the images approximately shift invariant. During the classification by VDM we only use the consistency of the rotations. Once we identify the nearest neighbors and rotational alignment, we search for shift alignment in the small neighborhood. Although we cannot globally center the projection images in class averaging step, the centers can be estimated later on using common-lines~\citet{Shkolnisky2012}.

\section{Experimental Results}
\label{sec:exp}
We performed numerical experiments to test the speed and accuracy of our algorithm on a machine with $2$ Intel (R) Xeon (R) CPUs X7542, each with $6$ cores, running at $2.67$ GHz with $256$ GB RAM in total. These experiments were performed in MATLAB in UNIX environment.

\subsection{Simulated noisy data}
\label{subsec:sim_noisy_data}
We compared our algorithms on simulated data against five 2D classification methods: RFA with K-means clustering implemented in SPIDER~\citet{SPIDERprotocol}, MSA/MRA implemented in IMAGIC~\citet{IMAGIC1996}, e2refine2d in EMAN2~\citet{Tang2007}, Relion 2D classification~\citet{Scheres2012}, and Xmipp CL2D~\citet{Sorzano2010}. The volume of \textit{E. coli} 70S ribosome-elongation factor G (EF-G)~\citet{SPIDERprotocol} was used to simulate projections. The image size is $129 \times 129$ pixels with $ 2.82 \unit{\AA}$/pixel. 
\begin{figure}[h]
\begin{center}
\subfloat[Clean]{
\includegraphics[width=0.22\textwidth]{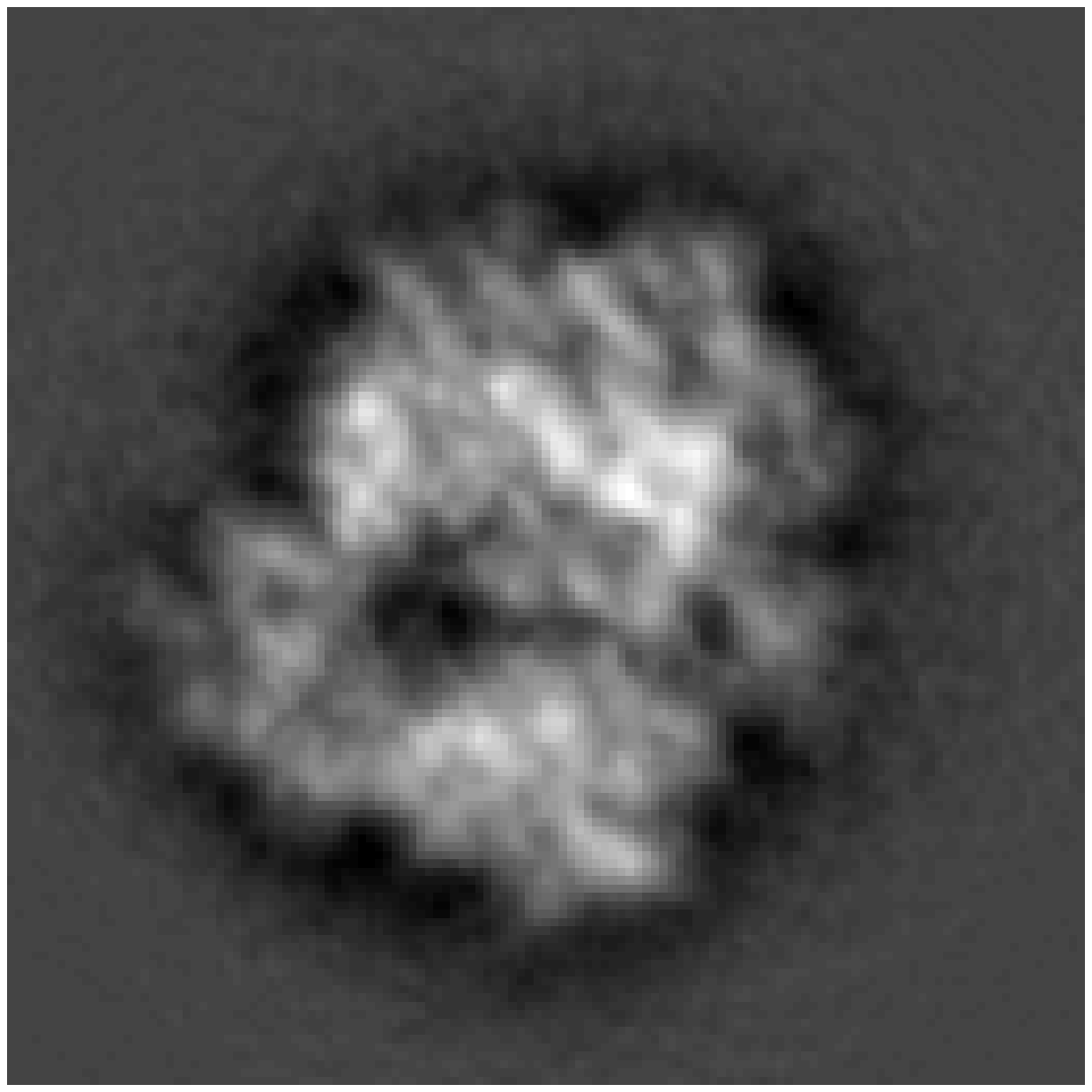} %
\label{fig:simulate_clean}
}
\subfloat[CTF modified]{
\includegraphics[width=0.22\textwidth]{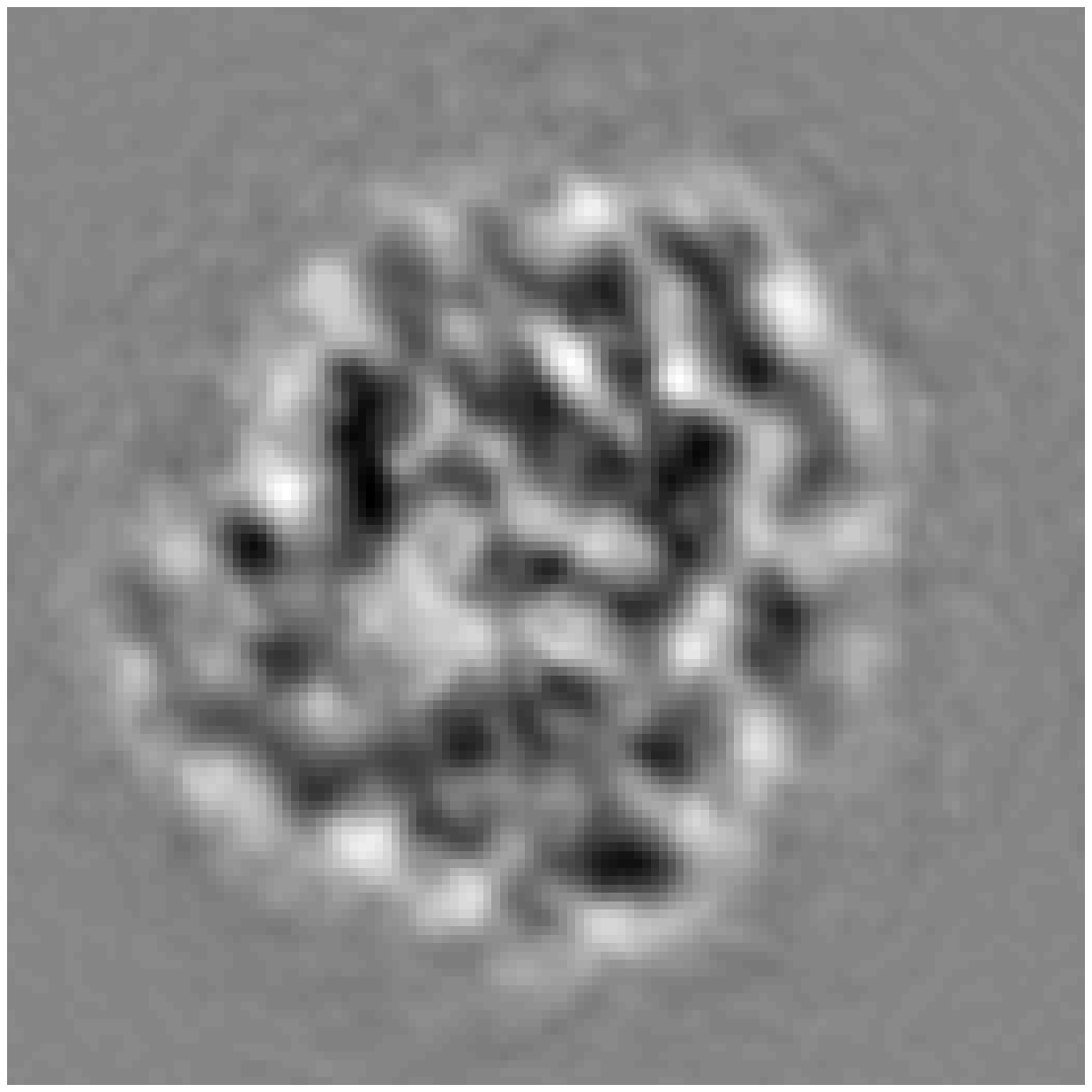} %
\label{fig:simulate_wctf}
}
\subfloat[SNR$=1/50$]{
\includegraphics[width=0.22\textwidth]{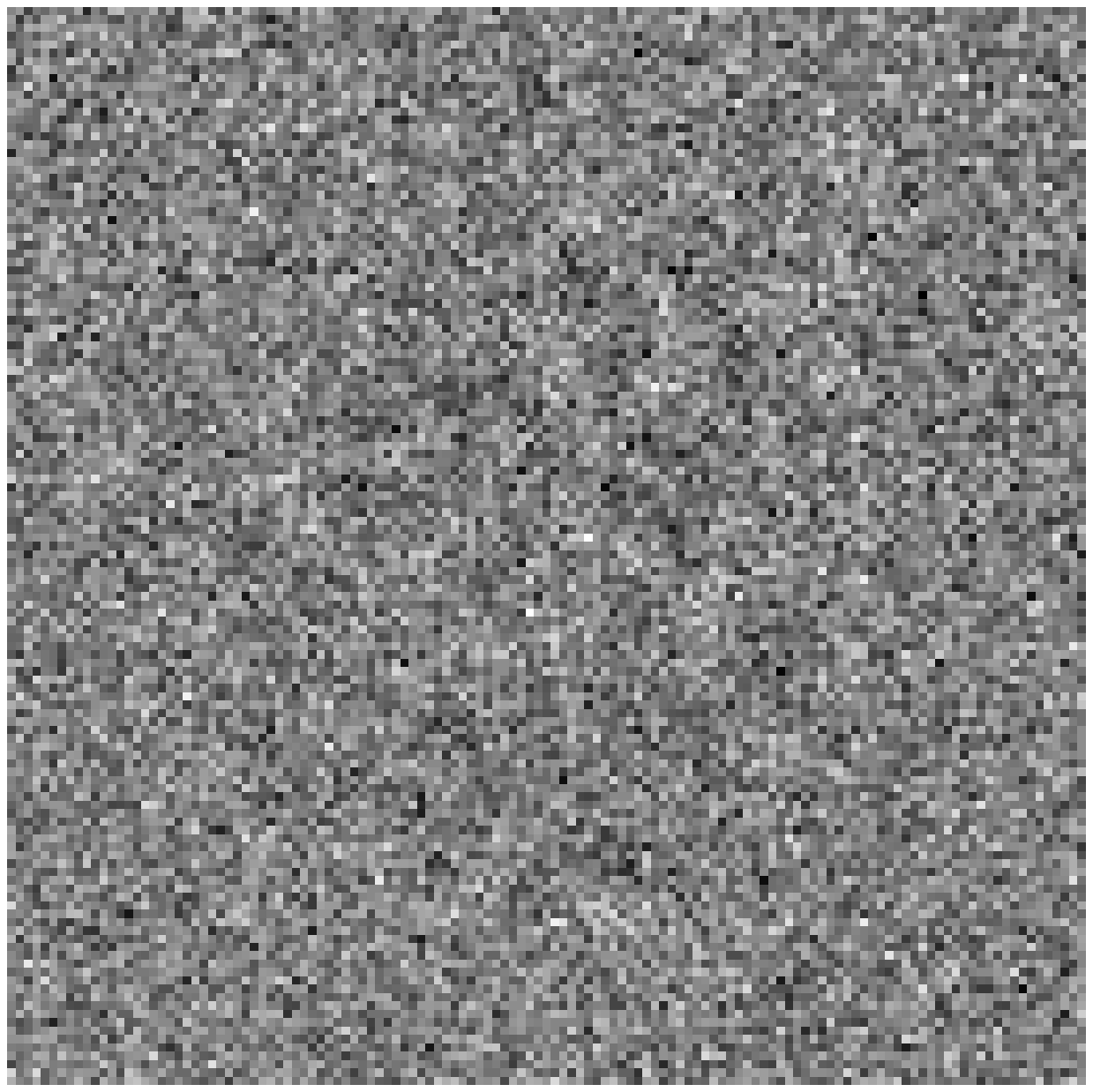} %
\label{fig:simulate_nsr50}
}\\
\subfloat[SNR$=1/100$]{
\includegraphics[width=0.22\textwidth]{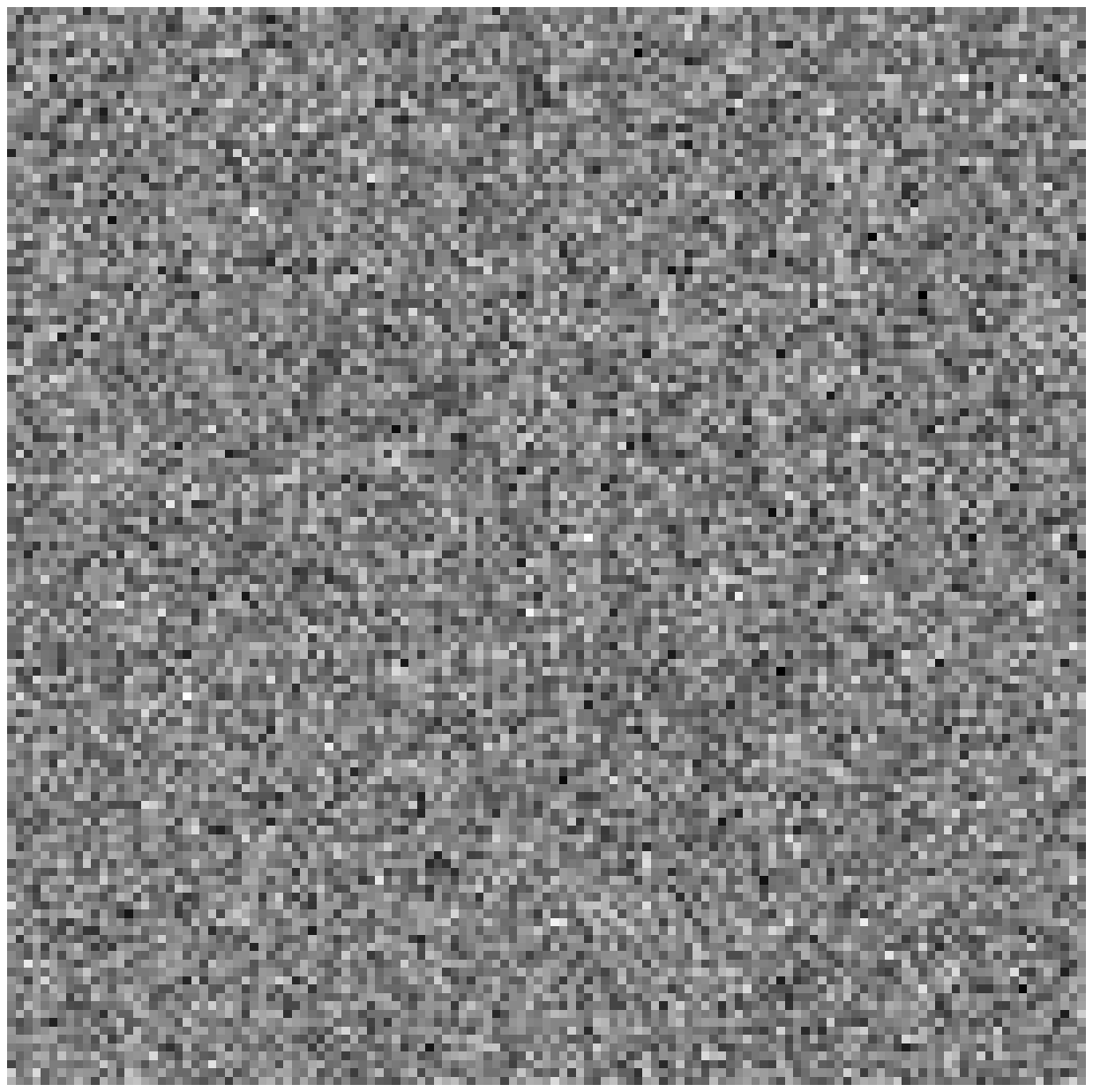} %
\label{fig:simulate_nsr100}
}
\subfloat[SNR$=1/150$]{
\includegraphics[width=0.22\textwidth]{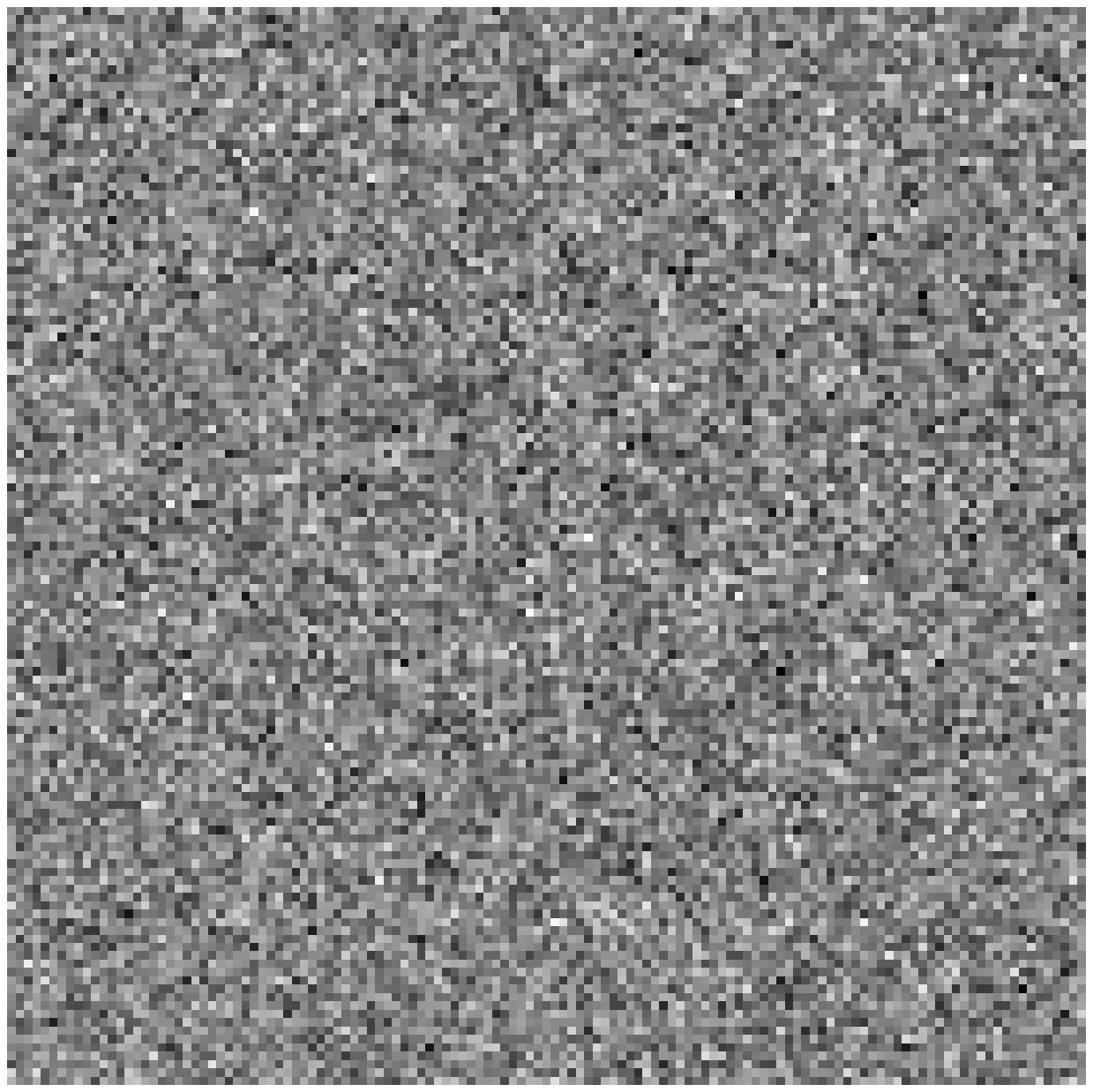} %
\label{fig:simulate_nsr150}
}
\subfloat[SNR$=1/200$]{
\includegraphics[width=0.22\textwidth]{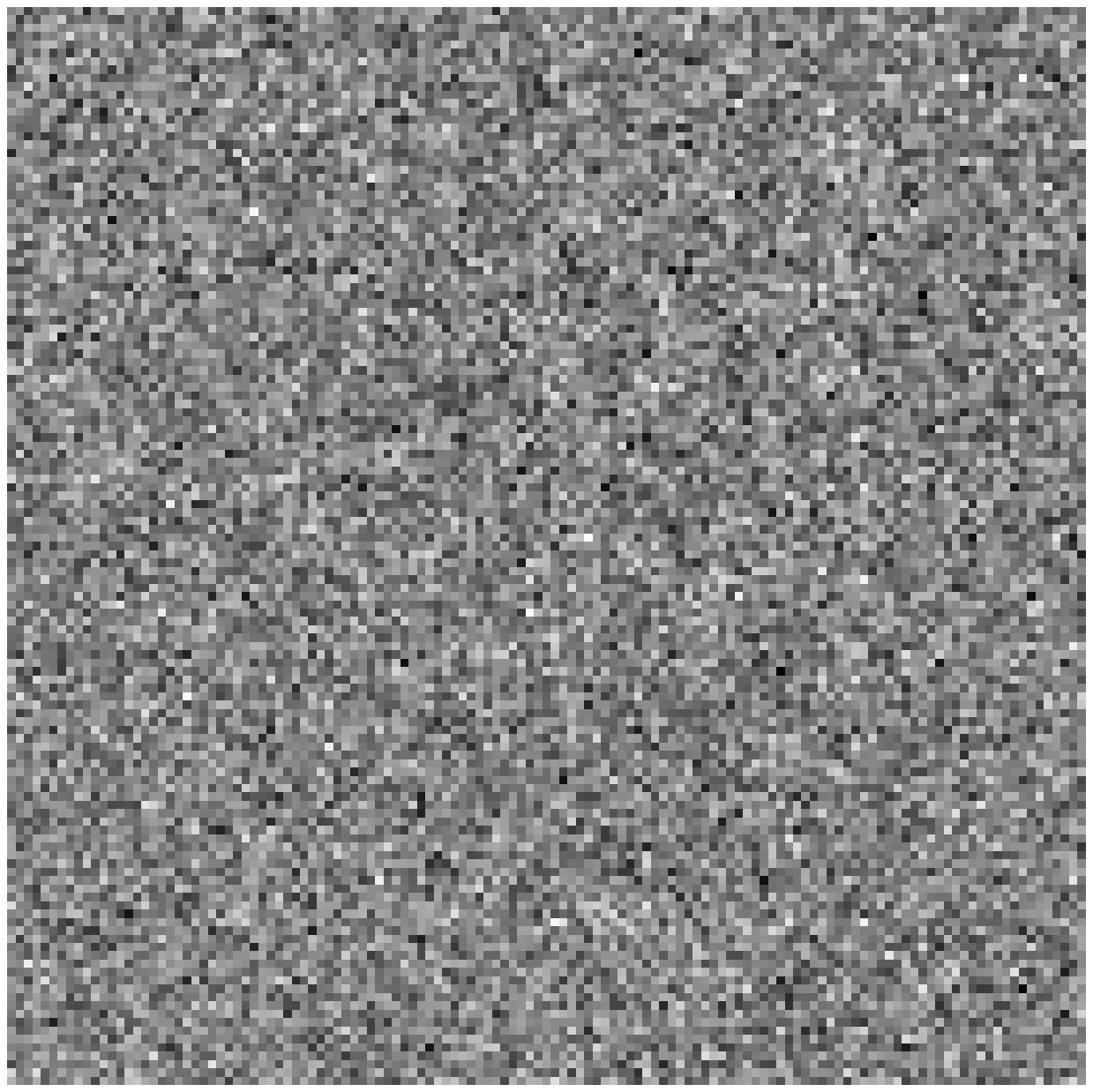} %
\label{fig:simulate_nsr200}
}
\end{center}
\caption{Simulated 70S ribosome projection images. (a) Simulated clean centered projection image. (b) Clean projection image modified by Gaussian envelope function and Contrast Transfer Function (CTF). (c), (d), (e), and (f) Slightly shifted (randomly shifted within the range of $\pm 4$ pixels in $x$ and $y$ directions) projection images with CTF contaminated with white Gaussian noise at SNR=$1/50, 1/100, 1/150,\text{ and }1/200$.} \label{fig:70S_noisyProj_wCTF}
\end{figure}
Images observed by an electron microscope are not true projections of the specimen. Imaging modifications include the effects of the contrast transfer function (CTF), which is introduced through electron lens aberrations and defocusing~\citet{Zhu1997}, and also the envelope function of the microscope, which contains contributions from a number of effects, such as spatial and temporal coherence, specimen motion, etc.~\citet{Hanszen1971}. In addition, background noise is present from a variety of sources. Therefore, we attempted to closely emulate the image formation process in the electron microscope including the effects of CTF, envelope function and noise. We projected $10^4$ clean images at directions sampled uniformly over the sphere (see Figure~\ref{fig:simulate_clean}). Then a Gaussian low-pass filter with half-width $1/10 \unit{\AA^{-1}}$ was applied to simulate the effect of the envelope function. CTFs with different defocus values were applied to the images (see Figure~\ref{fig:simulate_wctf}).
The contrast transfer functions are generated according to the formula, 
\begin{equation}
\text{CTF}(f) = \sin(\pi \lambda f^2(\Delta z - 0.5 \lambda^2 f^2 c_s)) + B \cos(\pi \lambda f^2(\Delta z - 0.5 \lambda^2 f^2 c_s)),
\end{equation}
where the variable $f$ is the spatial frequency, $\Delta z$ is the defocus, $c_s$ is the spherical abberation, $\lambda$ is the electron wavelength, and $B$ is the fraction of amplitude contrast. The imaging parameters were taken from the simulative data in SPIDER protocol~\citet{SPIDERprotocol}: electron beam energy $ E=200 \unit{KeV}$ with wavelength $\lambda= 0.025 \text{\AA}$ and spherical abberation is $c_s = 2.26 \unit{mm}$. The images were divided into $20$ different defocus groups, with minimum defocus $1.5 \unit{\mu m}$ and maximum defocus $4 \unit{\mu m}$. 

The centered projection images are randomly shifted within the range of $\pm 4$ pixels in $x$ and $y$ directions. The images are then contaminated with additive white Gaussian noise at different signal to noise ratios, SNR$=1/50, 1/100, 1/150,$ and $1/200$ (see Figure~\ref{fig:70S_noisyProj_wCTF}). The SNR in all our experiments is defined by
\begin{equation}\label{eq:snr definition}
\text{SNR} = \frac{\operatorname{Var}(\text{\textit{Signal}})}{\operatorname{Var}(\text{\textit{Noise}})}.
\end{equation}
The input images to our algorithm are first CTF corrected by phase flipping. More sophisticated CTF corrections are possible, but we find that phase flipping already produces satisfactory results.
\begin{table}
\begin{center}
{\footnotesize
    \begin{tabular}{| c | c | c | c | c | c | c |}
    \hline
                & RFA/K-means & MSA/MRA  & Relion & EMAN2 & Xmipp & ASPIRE \\ \hline
    SNR$=1/50$  & 0.45  & 0.97  & 0.79  & 0.74 & 0.83  & \textbf{1.00} \\ \hline
    SNR$=1/100$ & 0.09  & 0.87  & 0.70  & 0.45 & 0.68  & \textbf{0.99} \\ \hline
    SNR$=1/150$ & 0.07  & 0.67  & 0.52  & 0.13 & 0.48  & \textbf{0.90} \\ \hline \hline
    Timing (hrs) & 1.5 & 7.5 & 16 & 12 & 42 & \textbf{0.5} \\ \hline
\end{tabular}
}
\end{center}
\caption{Proportion of viewing angles of nearest neighbors that lie within $18.2^\circ$. Experiments are performed with $10^4$ projection images of 70S ribosome at different noise levels. RFA was performed with AP SR program in SPIDER, the aligned particles were then classified into 200 groups using K-means algorithm. MSA/MRA was implemented in IMAGIC and was iterated for 5 times. We performed 25 iterations of Relion 2D class averaging, 10 iterations of e2refine2d in EMAN2, and 60 iterations of CL2D in Xmipp. The particles were classified in 200 classes, so that on average there were 50 particles in each class. In our algorithm, we found $50$ nearest neighbors for each particle. The running time is measured for data with SNR$=1/100$.}
\label{tab:sim_res_1}
\end{table}

In our simulation we know the original viewing angles, so for each image we compute the angles (in degrees) between the viewing angle of the image and the viewing angles of its $50$ nearest neighbors. Small angles indicate successful identification of ``true'' neighbors that belong to a small spherical cap, while large angles correspond to outliers. We compute the percentage of nearest neighbor pairs whose viewing angles are within $18.2^\circ$ spherical cap ($\cos (18.2^\circ) = 0.95$) as a measure of the quality of 2D image classification (see Table~\ref{tab:sim_res_1}).  

For experiments performed in SPIDER, all phase-flipped noisy images were filtered with a low-pass Butterworth filter, with the pass band and stop band at $0.08$ and $0.12$ respectively, given in reciprocal pixels, as described in~\citet{SPIDERprotocol}. To convert these values to Angstroms, divide the pixel size by the spatial frequency, i.e., in our case, $2.82/0.12 \text{\AA}^{-1} = 23.5 \text{\AA}$. We used a program in SPIDER (AP SR) to perform RFA on band-pass filtered projection images. K-means clustering was used to classify the aligned and filtered images into $K=200$ groups. Software description and details for performing the 2D image classification in SPIDER are available in~\citet{SPIDERprotocol}. The running time for generating 200 class averages is 1.5 hours (see Table~\ref{tab:sim_res_1}).

\begin{figure}[h!]
\begin{center}
\subfloat[IMAGIC]{
\includegraphics[width=0.3\textwidth]{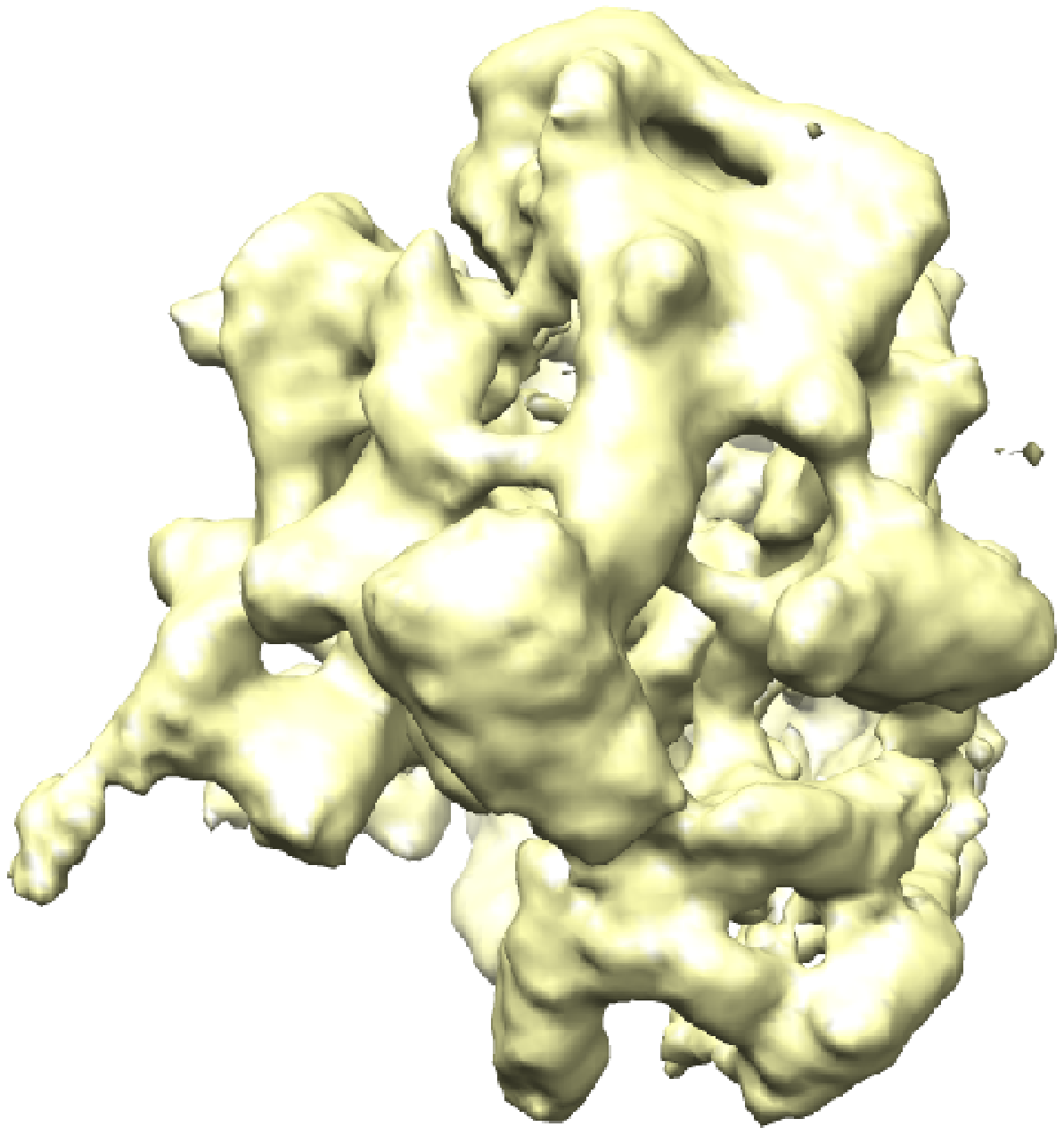}%
\label{fig:70S_IMAGIC}
}
\subfloat[Relion]{
\includegraphics[width=0.3\textwidth]{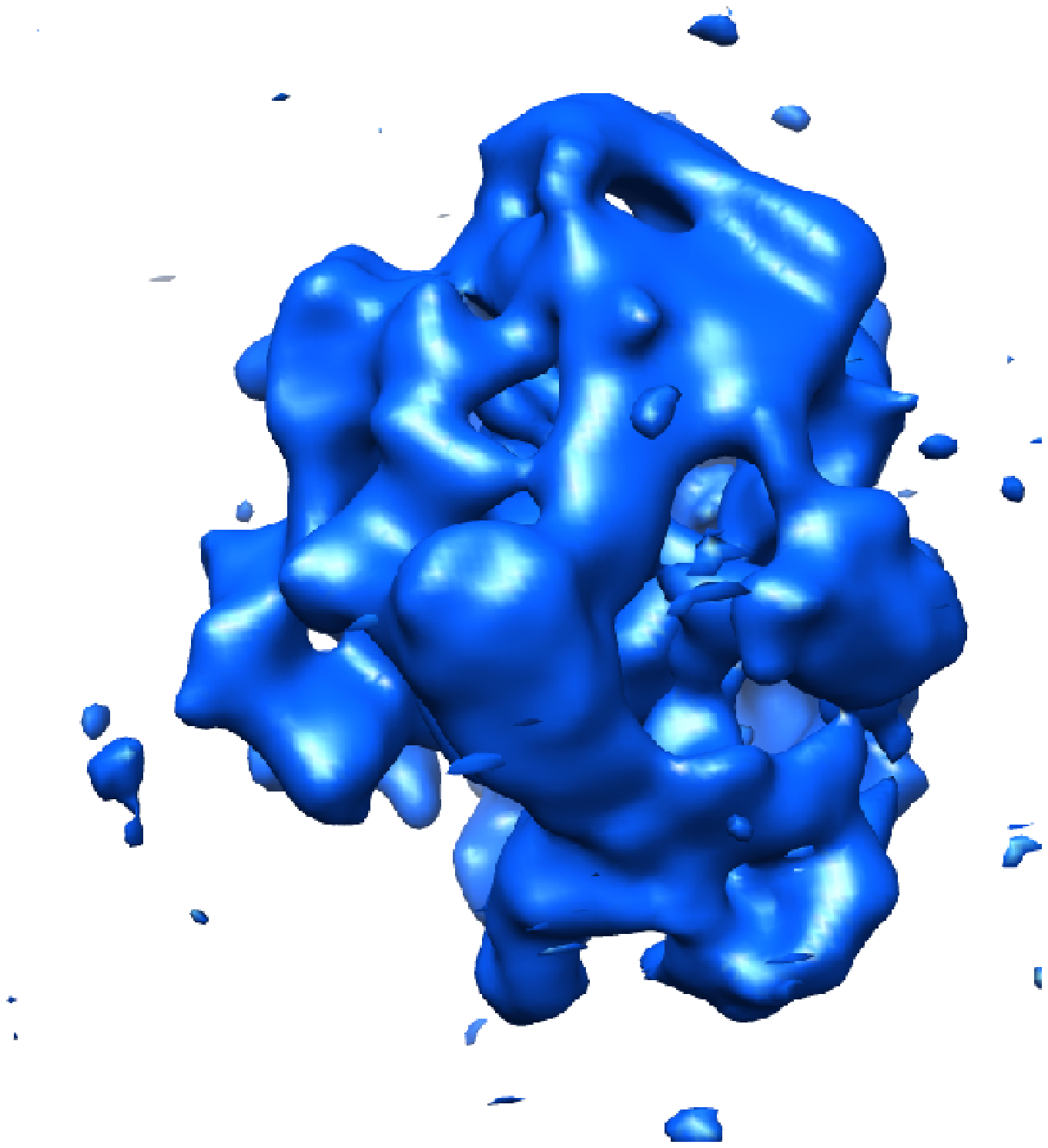}%
\label{fig:70S_Relion}
}
\subfloat[Xmipp]{
\includegraphics[width=0.3\textwidth]{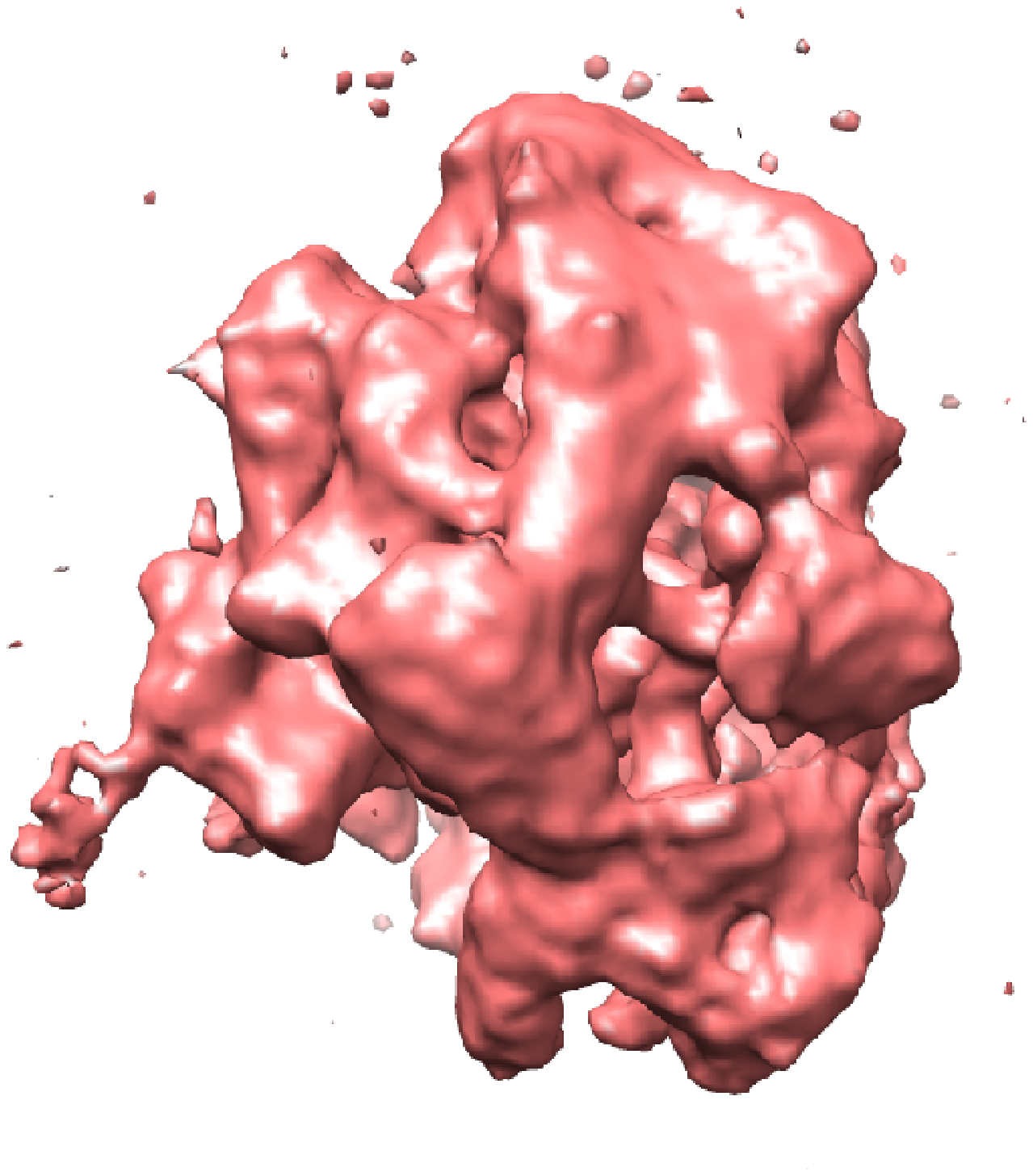}%
\label{fig:70S_Xmipp}
}\\
\subfloat[EMAN2]{
\includegraphics[width=0.3\textwidth]{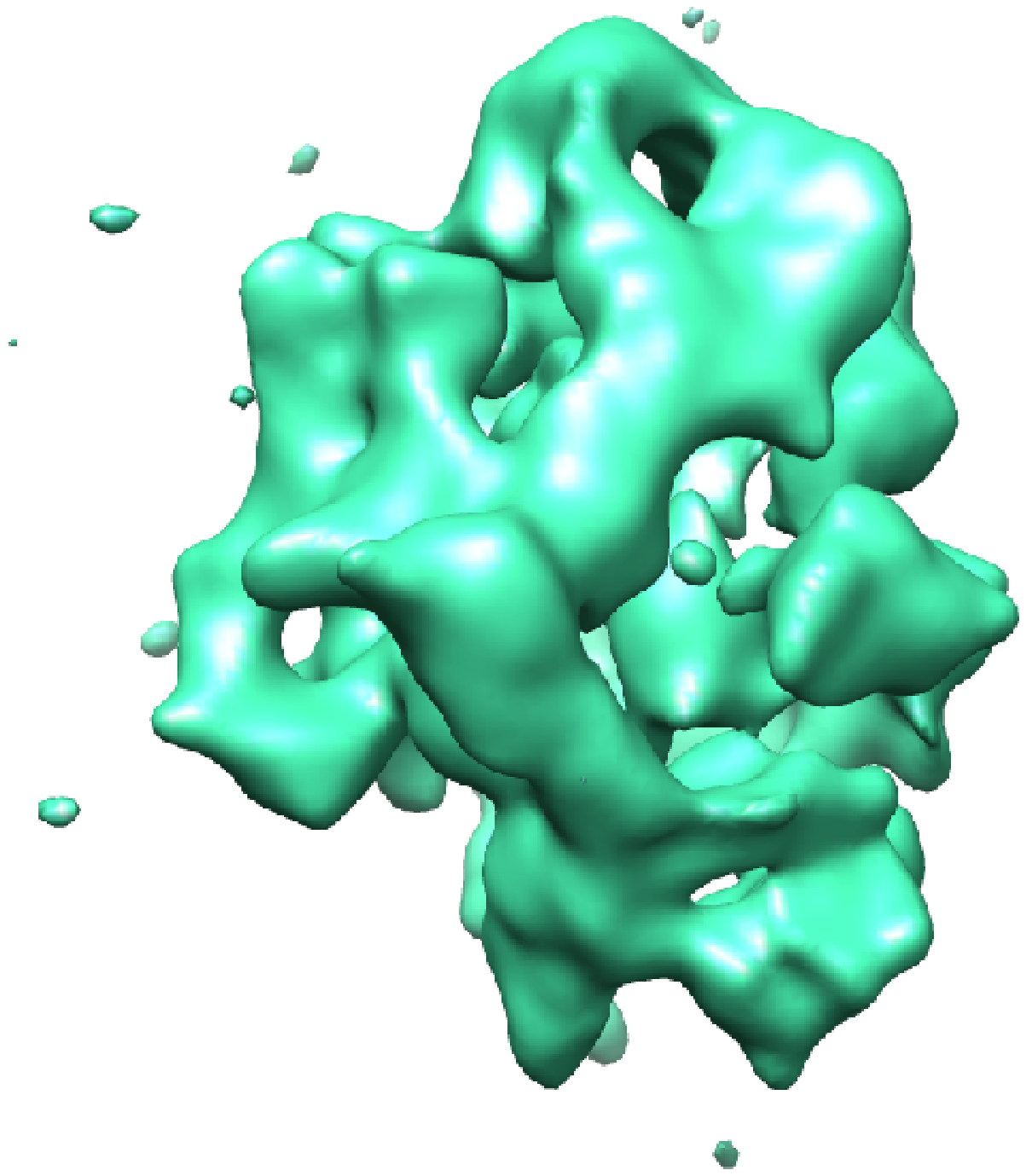}%
\label{fig:70S_EMAN2}
}
\subfloat[ASPIRE]{
\includegraphics[width=0.3\textwidth]{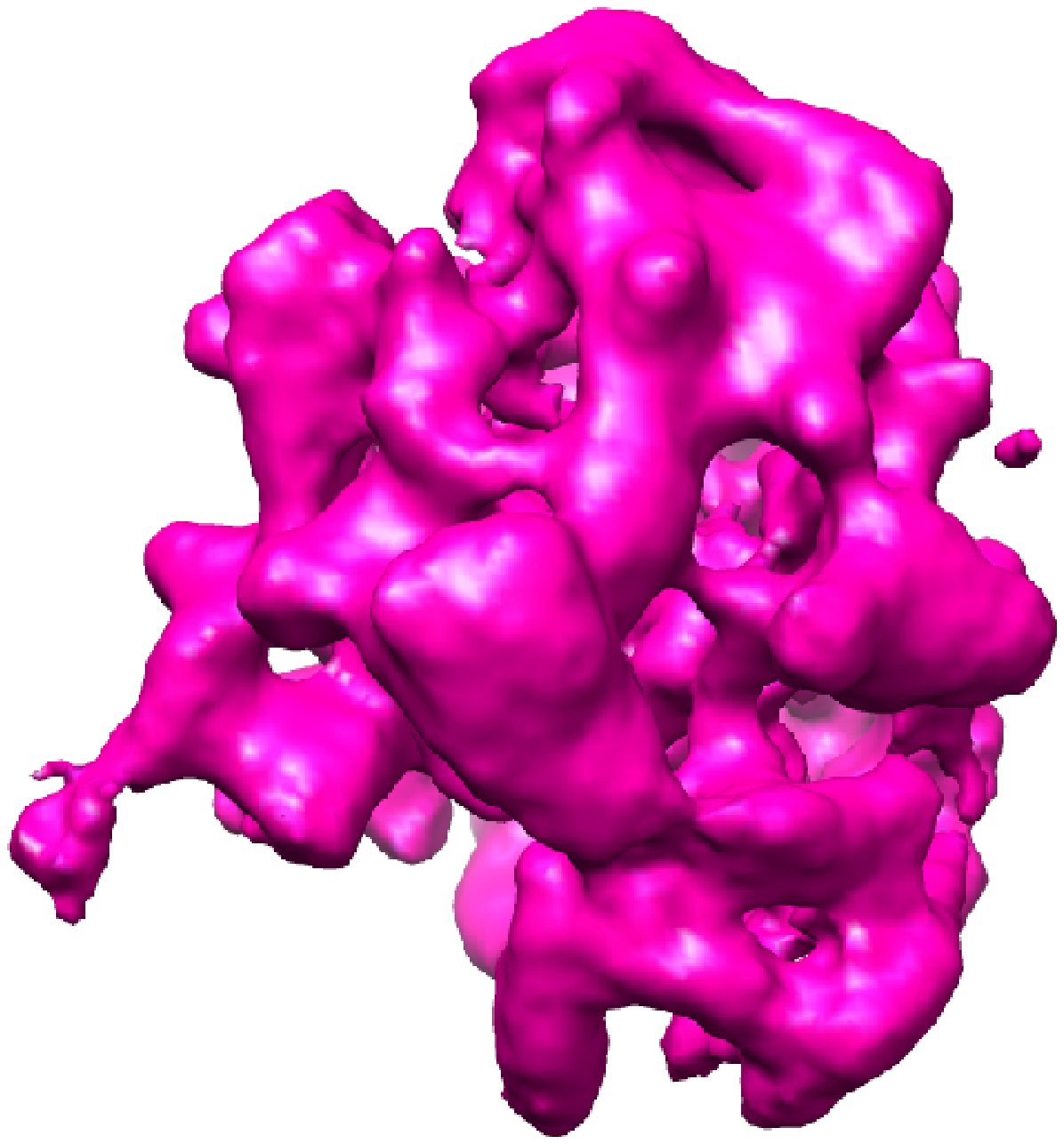}%
\label{fig:70S_aspire}
}
\subfloat[Reference]{
\includegraphics[width=0.3\textwidth]{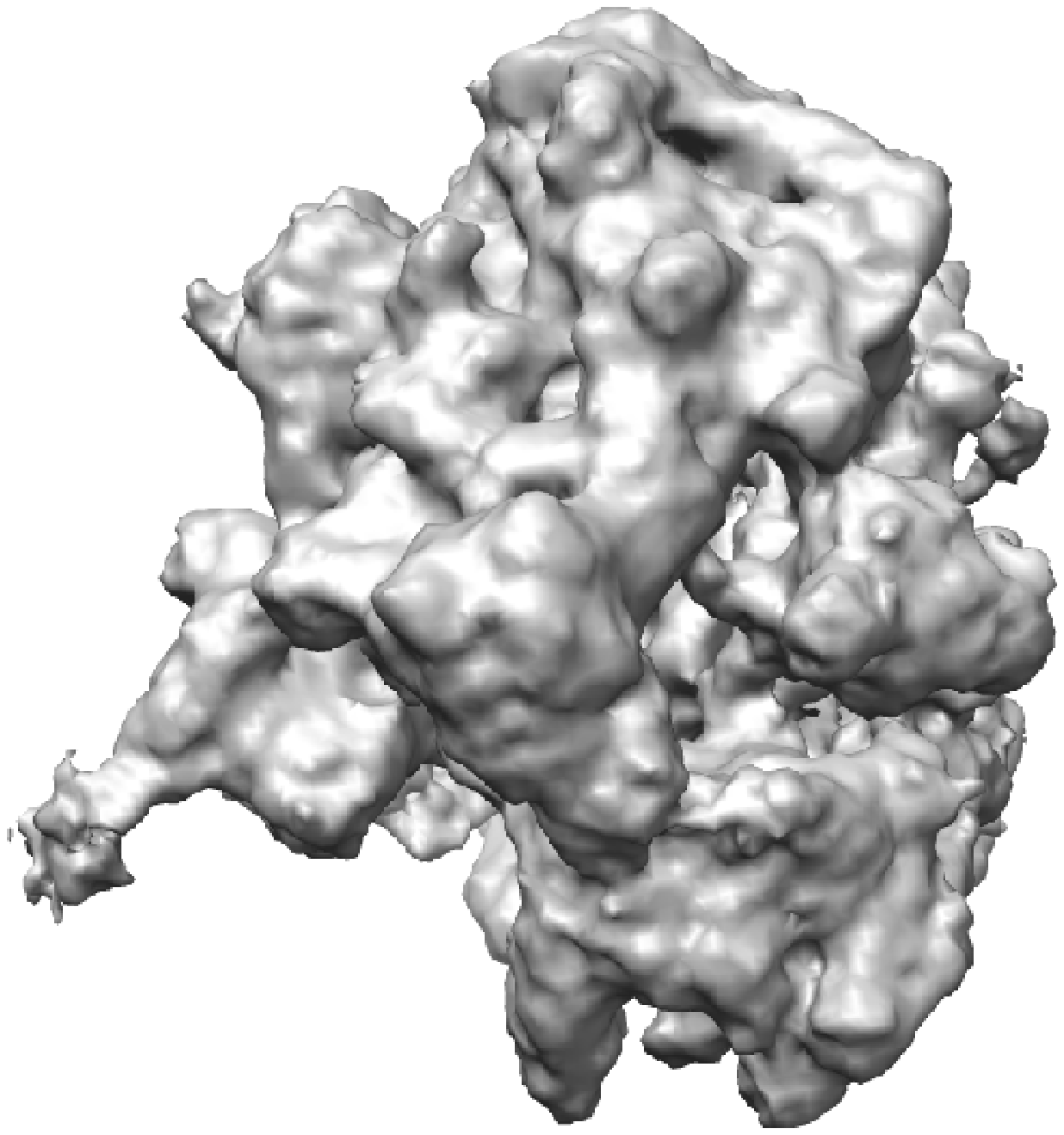}%
\label{fig:ref}
}
\end{center}
\caption{\textit{Ab initio} models of 70S obtained from $10^4$ simulated noisy projection images (SNR=1/100) with $20$ defocus groups. The ab initio models are obtained by assigning orientations to the class averages using the common-lines based LUD  method~\citet{Wang2013}. Reconstructed volumes from class averages generated by (a) MSA/MRA 2D image classification implemented in IMAGIC with 5 iterations, (b) 2D class averaging in Relion with 25 iterations, (c) CL2D in Xmipp with 60 iterations, (d) e2refine2d in EMAN2 with 10 iterations, and (e) 2D class averaging in ASPIRE (described in this paper). (f) Reference volume. The reconstructed volumes are Gaussian filtered.}
\label{fig:70S_volumes}
\end{figure}

For the experiments performed in IMAGIC, images were crudely centered by correlating the images with the data mean iteratively. The crudely centered images were first classified into 50 classes using MSA. Then 50 reference images were generated and the projection images were aligned with the references using multi-reference alignment. The aligned images were classified into 200 groups. The multi-reference alignment and MSA classification into 200 classes were iterated 3 more times to get the final alignment and classification results. More iterations of the MSA/MRA classification can improve the classification result. However each iteration took about 2 hours to finish for this data set. 

We also tested the more modern cryo-EM SPR packages EMAN2, Xmipp, and Relion. The program e2refine2d in EMAN2 is very similar to the MSA/MRA algorithm in IMAGIC. The difference is that the initial classification is done on translational and rotationally invariant features. We used 10 iterations of the 2D class averaging in EMAN2.   For experiments performed in Xmipp, we used CL2D algorithm for generating 2D class averages. The images were classified into 8 classes initially and then refined into, 16, 32, 64, 128, and finally 200 classes. In each level, there were 10 iterations to refine classification and alignment. Relion employs an empirical Bayesian approach for 2D classification. We ran 25 iterations of 2D class classification in Relion. The accuracy and running time for 2D classification are detailed in Table~\ref{tab:sim_res_1}. 

\begin{table}
\begin{center}
    \begin{tabular}{| c | c | }
    \hline
    \textbf{Step} & \textbf{Time (sec)} \\ \hline
    Fourier-Bessel sPCA  & $537.7$  \\ \hline
    Rotationally Invariant Features & $28.2$ \\ \hline
    Initial Nearest Neighbor Search & $13.9$ \\ \hline
    VDM Classification & $57.4$ \\ \hline
    Local Alignment and Class Average & $1081$ \\ \hline \hline
    \textbf{Total} & \textbf{1718.3 (28.6 min)} \\ \hline
    \end{tabular}
\end{center}
\caption{Timing for different steps of our 2D class averaging algorithm.}
\label{tab:timing}
\end{table}

We applied our rotational invariant viewing angle classification on the phase-flipped images. Our rotational invariant classification achieves better classification results in finding particles of similar views than the other five methods (see Table~\ref{tab:sim_res_1}). Each image was aligned and averaged with its $50$ nearest neighbors. It took about half an hour to generate $10^4$ class averages. Table~\ref{tab:timing} summarizes the timing for each step of our algorithm. 

In another set of experiments, we used Fourier-Bessel steerable PCA denoised images (SNR$=1/100$) as the input for both SPIDER, IMAGIC, EMAN2 and Xmipp 2D classification programs. The classification results are greatly improved (see Table~\ref{tab:sim_res_2}). This demonstrates that the denoising scheme we used in our pipeline is very useful for 2D image classification.  
\begin{table}
\begin{center}
    \begin{tabular}{| c | c | c | }
    \hline
      & no FBsPCA denoising & FBsPCA denoising\\ \hline
    RFA/K-means & 0.09  & \textbf{0.48} \\ \hline
    MSA/MRA & 0.87 & \textbf{0.95} \\ \hline
    EMAN2 & 0.45 & \textbf{0.76} \\ \hline
    Xmipp & 0.68 & \textbf{0.96} \\
    \hline
    \end{tabular}
\end{center}
\caption{Denoising using FBsPCA improves the classification results in RFA/K-means, MSA/MRA, EMAN2 and Xmipp 2D image classification (SNR$=1/100$). Values in the table are the proportion of the viewing angles of particles in the same class that are within $18.2^\circ$.}
\label{tab:sim_res_2}
\end{table}

\begin{figure}
\begin{center}
\includegraphics[width=0.8\textwidth]{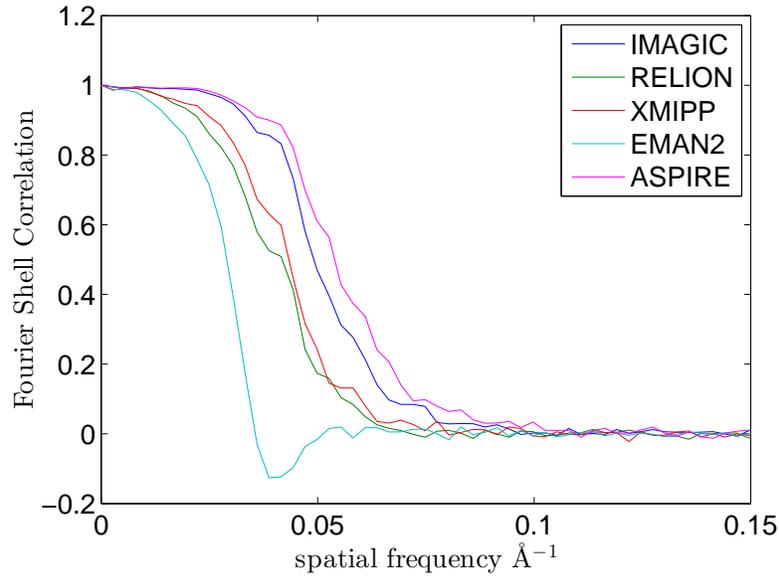}
\end{center}
\caption{Fourier shell correlation of the reference volume with the ab initio models from different class averages (IMAGIC, Relion, Xmipp, EMAN2, and ASPIRE).}
\label{fig:FSCs}
\end{figure}

The resulting class averages were used to find common-lines. An \textit{ab initio} estimate of the 3D orientations was determined by the least unsquared deviation (LUD) method~\citet{Wang2013}, which is also available in the ASPIRE toolbox under ``est$\_$orientations$\_$LUD.m''. The reconstructed volumes from the class averages are shown in Figure~\ref{fig:70S_volumes}. We were unable to reconstruct a meaningful model from the class averages generated by RFA/K-means procedure due to the large error in classification. The reconstructed volumes from the class averages produced by IMAGIC, Relion, Xmipp and EMAN2 and this paper were compared with the reference volume (Figure~\ref{fig:ref}). The {\it ab initio} model built from the class averages with this paper's methods agrees best with the reference volume (see Figure~\ref{fig:FSCs}).

\begin{figure}
\begin{center}
\includegraphics[width=0.8\textwidth]{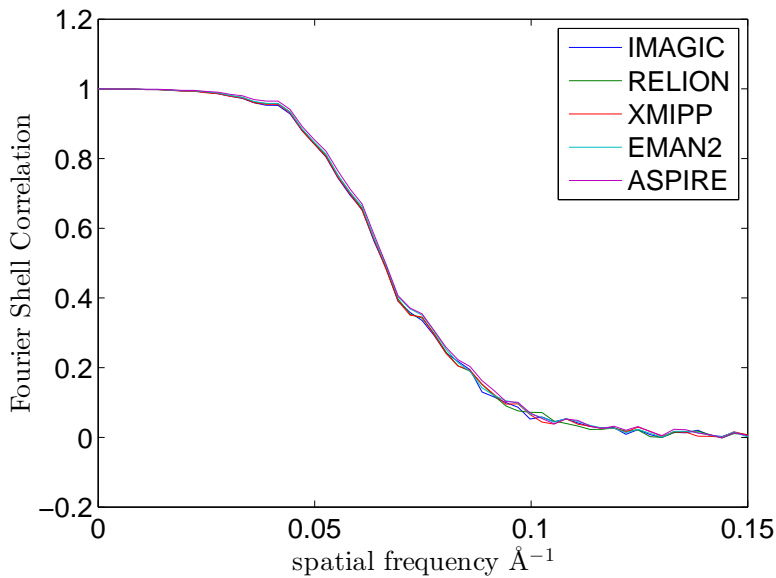}
\end{center}
\caption{Fourier shell correlation of the reference volume with the refined models from different ab initio models (IMAGIC, Relion, Xmipp, EMAN2, and ASPIRE).}
\label{fig:FSCs_refined}
\end{figure}

\begin{table}
\begin{center}
    \begin{tabular}{| c | c | c | c | c |}
    \hline
    IMAGIC  & Relion & EMAN2 & Xmipp & ASPIRE \\ \hline
     17 & 18  & 20 & 18 & \textbf{14} \\
    \hline
    \end{tabular}
\end{center}
\caption{Number of refinement iterations needed for convergence starting from different ab initio models (IMAGIC, Relion, Xmipp, EMAN, and ASPIRE). We used Relion 3D auto-refine for refinement.}
\label{tab:refine_iter}
\end{table}

After \textit{ab initio} reconstruction, we used Relion 3D auto-refine~\citet{RELIONtutorial} to refine those five different \textit{ab initio} models (IMAGIC, Relion, Xmipp, EMAN2, and ASPIRE) with simulated projection images whose SNR is $1/100$. The FSC curves look very similar for the refined models (see Figure~\ref{fig:FSCs_refined}). However it takes different number of iterations to reach convergence (see Table~\ref{tab:refine_iter}). Refinement starting from ASPIRE \textit{ab initio} model converged most quickly and it took $14$ iterations. The FSC curves (in Figure~\ref{fig:FSCs}) and the number of iterations (in Table~\ref{tab:refine_iter}) show that the quality of the \textit{ab inito} volume affects the refinement's convergence rate. 

\subsection{Experimental data: 70S ribosome}
\begin{figure}[h]
\begin{center}
\includegraphics[width=0.8\textwidth]{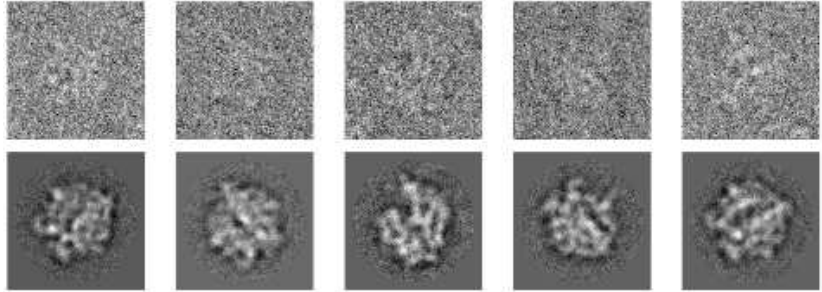} 
\end{center}
\caption[70S ribosome: raw images and class averages.]{Top row: Samples of experimental images for 70S ribosome. Bottom row: Class averages by averaging the raw images of the top row with their $50$ aligned nearest neighbors. Courtesy of Dr. Joachim Frank.} \label{fig:FrankData}
\end{figure}

We applied the pipeline of image denoising, classification and alignment to an experimental data set provided by Dr. Joachim Frank's group~\citet{Agirrezabala2012}. This data set comes from a larger heterogeneous data set with $216,517$ particles. ML3D~\citet{Scheres2007} was used to separate the data into $6$ more homogeneous subsets. The data used here is class number $6$ and contains $40,778$ projection images of 70S ribosome (see top row of Figure \ref{fig:FrankData}). The images are of size $250\times 250$ pixels with $1.5\text{\AA}$/pixel and the electron beam wavelength $\lambda=0.0197\text{\AA}$. They were pooled together from $77$ different defocus groups and CTF corrected by phase-flipping. We split the data set randomly into two equally sized groups, each containing $20,389$ images. $50$ nearest neighbors and the corresponding rotational and shift alignment were identified for each image. The second row of Figure~\ref{fig:FrankData} shows the averaged images.
$1500$ class averages were used to build a \textit{ab initio} model for each group, with the common-lines based method~\citet{SingerSchkolnisky, Wang2013} for orientation determination.
\begin{figure}
\begin{center}
\subfloat[Recon. 1]{
\includegraphics[width=0.3\textwidth]{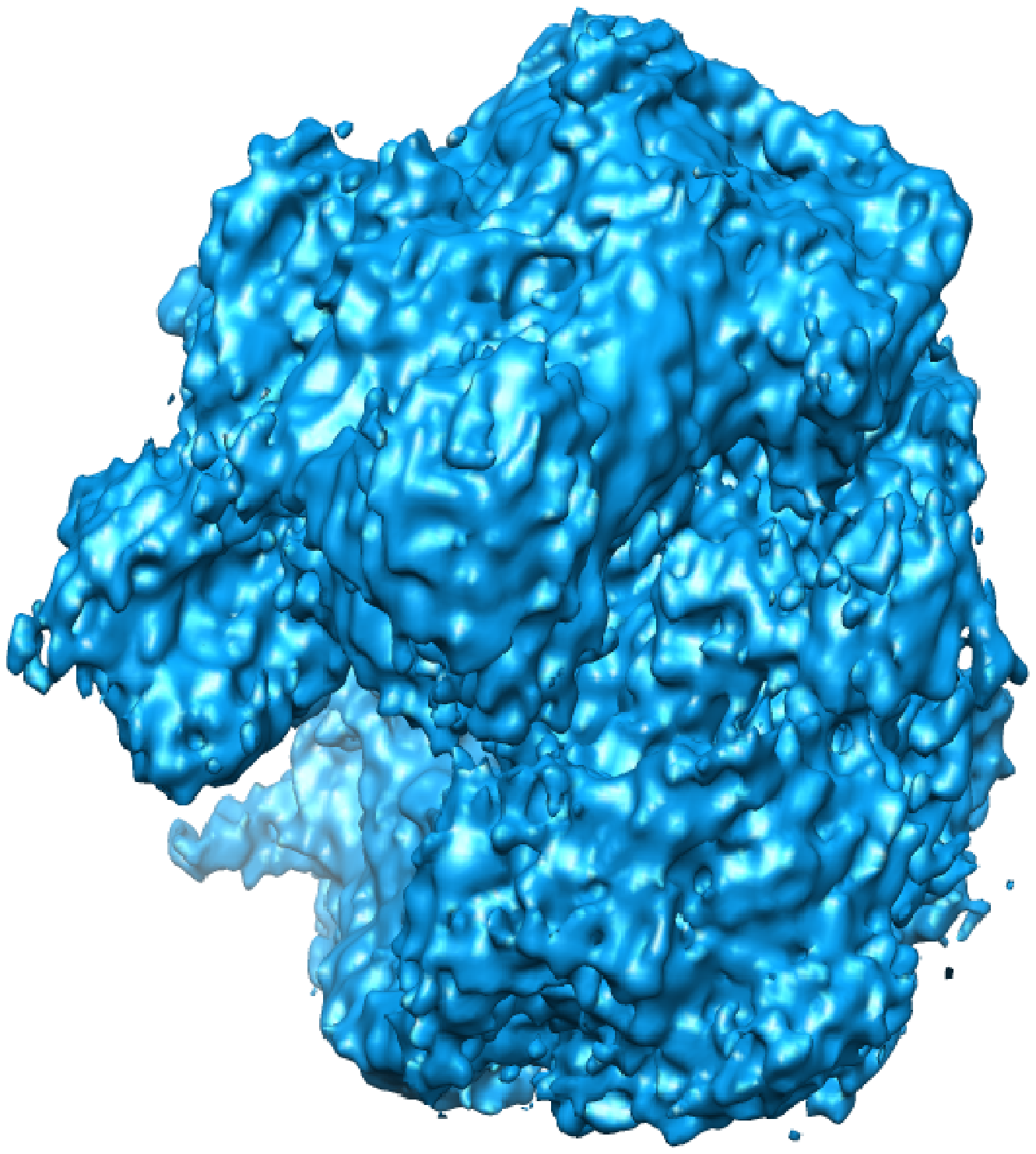} %
\label{fig:FrankData_vol1}
}
\subfloat[Recon. 2]{
\includegraphics[width=0.3\textwidth]{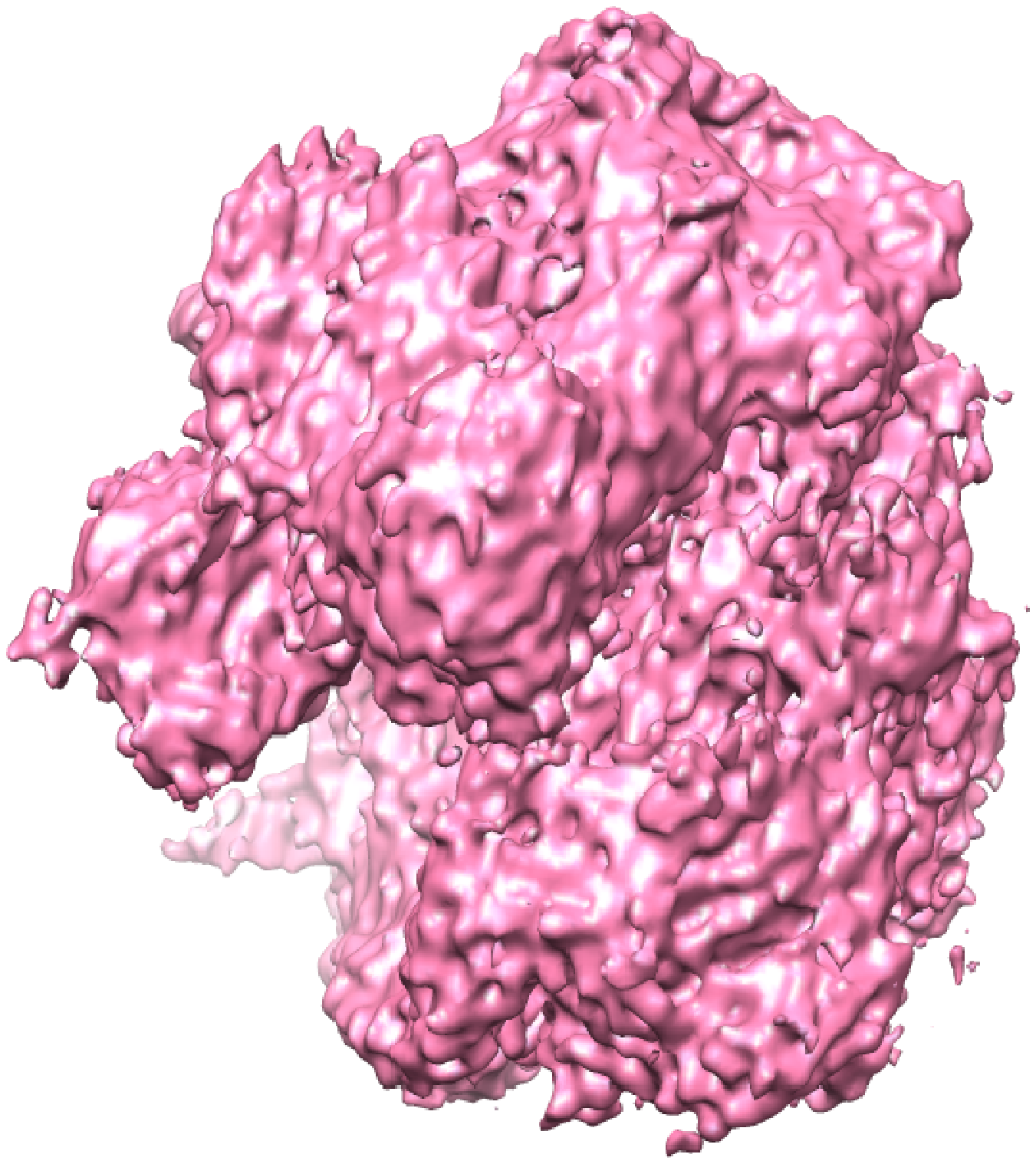} %
\label{fig:FrankData_vol2}
}
\end{center}
\caption{\textit{Ab initio} reconstructions of 70S ribosome from two independent data sets. (a) Snapshot of ab initio volume 1. (b) Snapshot of ab initio volume 2.} \label{fig:FrankData_volumes}
\end{figure}

\begin{figure}
\begin{center}
\includegraphics[width=0.7\textwidth]{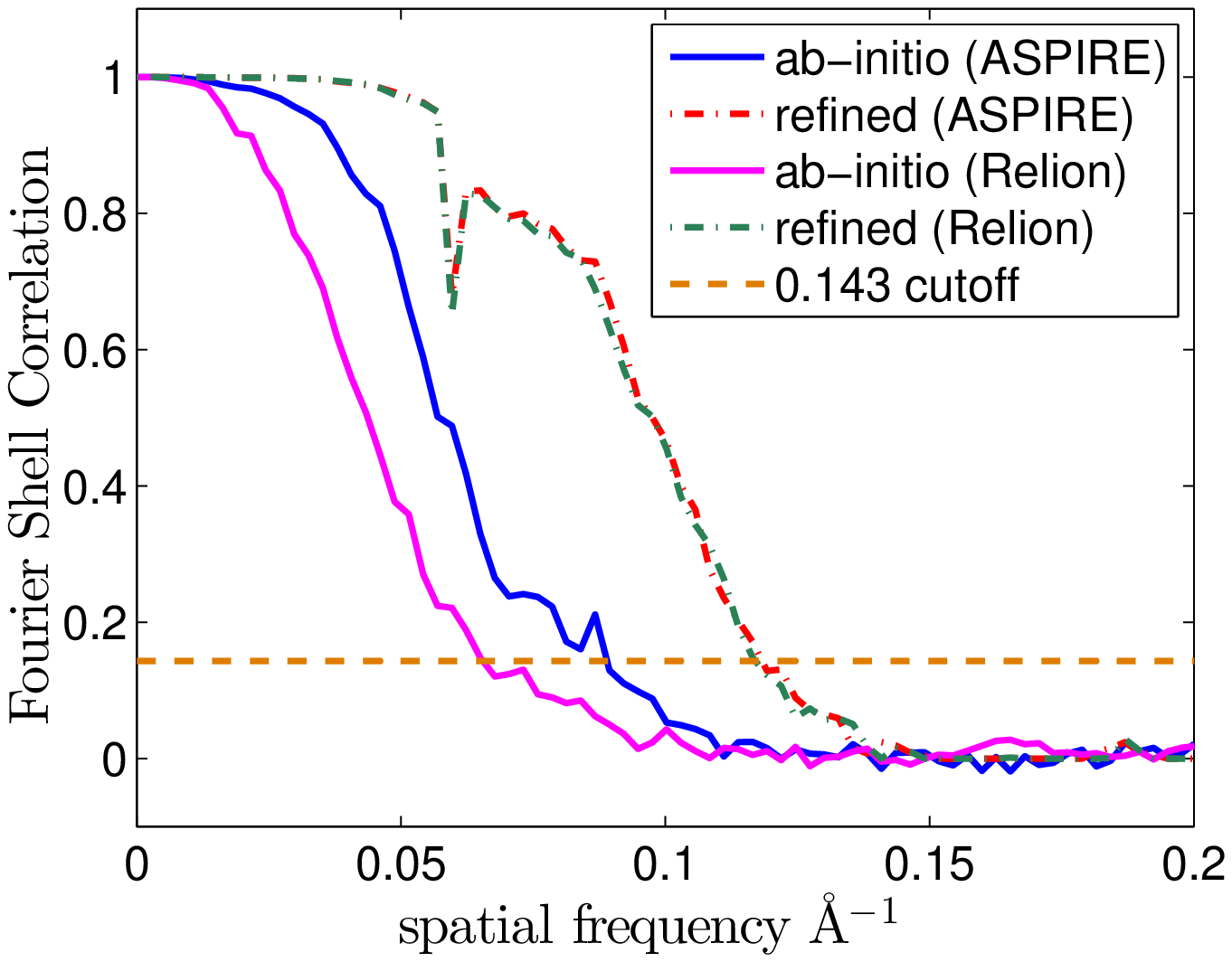} %
\end{center}
\caption{Fourier shell correlation curves for ab initio models and refined models. With $0.143$ cutoff criterion, the resolution is $11.53\text{\AA}$ for ASPIRE ab initio model (blue) and $15.38\text{\AA}$ for Relion ab initio model (magenta). Both refined models achieve $8.58\text{\AA}$ resolution according to gold-standard FSC (green and red dot-dash lines).}
\label{fig:70SFSC}
\end{figure}
 
The {\it ab initio} volumes (see Figure~\ref{fig:FrankData_vol1} and~\ref{fig:FrankData_vol2}) are consistent with each other up to $11.53\text{\AA}$. Below the corresponding frequency, the Fourier shell correlation (blue line in Figure~\ref{fig:70SFSC}) between the two volumes is above $0.143$. The \textit{ab initio} model was refined in Relion 3D auto-refine~\citet{RELIONtutorial}. The refined model achieves $8.58\text{\AA}$ resolution with 0.143 cutoff and $10.25\text{\AA}$ with 0.5 cutoff (see red dot-dash line in Figure~\ref{fig:70SFSC}). Our refined model achieves higher resolution than the previously reported resolution $11.5\text{\AA}$, with $0.5$ cutoff criterion for FSC~\citet{Agirrezabala2012}. Note that in our refinement process, two volumes were refined independently until the refinement converges whereas in the previous work~\citet{Agirrezabala2012}, the refinement was not done independently with the gold-standard FSC. With our \textit{ab initio} model, the refined model achieves higher resolution.

To compare with another 2D class averaging method, we used Relion 2D classification to generate $400$ class averages for each group. About $60$ good class averages in each group were chosen to generate \textit{ab initio} models. The resolution for the \textit{ab initio} model is $15.38\text{\AA}$ with $0.143$ cutoff criterion (see magenta line in Figure~\ref{fig:70SFSC}). The refined model achieves the same resolution as the refined model from ASPIRE (see Figure~\ref{fig:70SFSC}). The refinement took 20 iterations to converge, three more iterations than was needed for ASPIRE \textit{ab initio} model. Therefore, our 2D class averaging method improved the resolution of the \textit{ab initio} model of 70S ribosome and the refinement converged more quickly. 

\subsection{Experimental data: 50S ribosomal subunit}
\begin{figure}[h]
\begin{center}
\includegraphics[width = 0.8\textwidth]{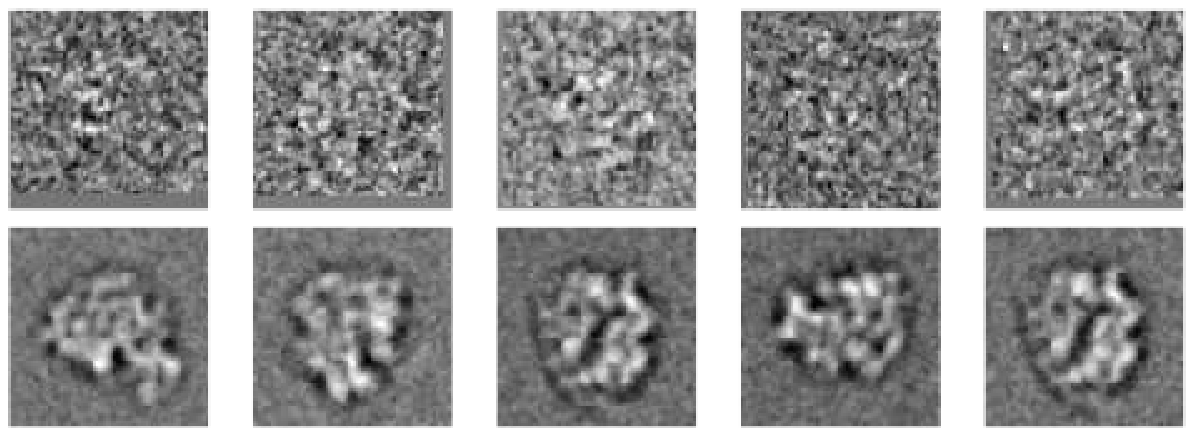} 
\end{center}
\caption[50S ribosomal subunit: raw images and class averages.]{Top row: Samples of experimental images of 50S ribosomal subunit. Bottom row: Class averages by averaging the raw images of the top row with their $50$ aligned nearest neighbors (including reflected images). Courtesy of Dr. Marin van Heel.} \label{fig:FredData}
\end{figure}

A set of micrographs of {\it E. coli} 50S ribosomal subunit was provided by Dr. Marin van Heel. We applied our algorithms to this data set, which contains $27,121$ projection images of the 50S ribosomal subunit. These micrographs were acquired by a Philips CM20 electron microscope at $9$ different defocus values between $1.37$ and $2.06 \unit{\mu m}$. Each image (see top row of Figure \ref{fig:FredData}) is of size $90 \times 90$ pixels with $3.36\text{\AA}$/pixel. The particles were picked using the automated particle picking algorithm in EMAN Boxer~\citet{Ludtke1999}. 
Then using the IMAGIC software package~\citet{IMAGIC1996}, the images were phase-flipped to remove the phase reversals in the CTF, bandpass filtered at $1/150$ and $1/8.4 \text{\AA}^{-1}$, and normalized by their variance. The images were initially crudely centered by correlating them with a fixed circularly-symmetric reference (rotationally averaged total sum of the data).

\begin{figure}
\begin{center}
\subfloat[Recon. 1]{
\includegraphics[width=0.3\textwidth]{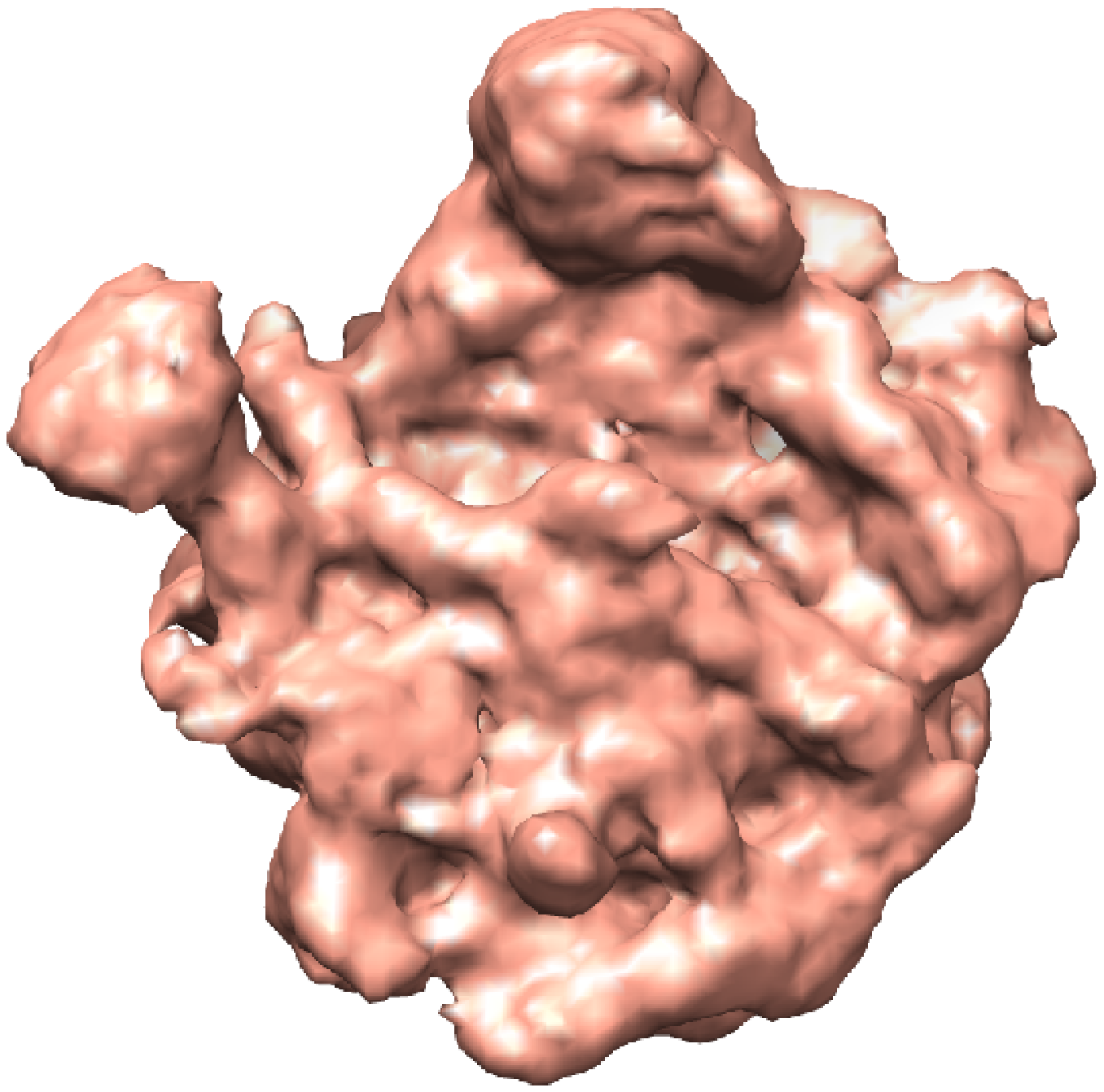}  %
\label{fig:FredData_vol1}
}
\subfloat[Recon. 2]{
\includegraphics[width=0.3\textwidth]{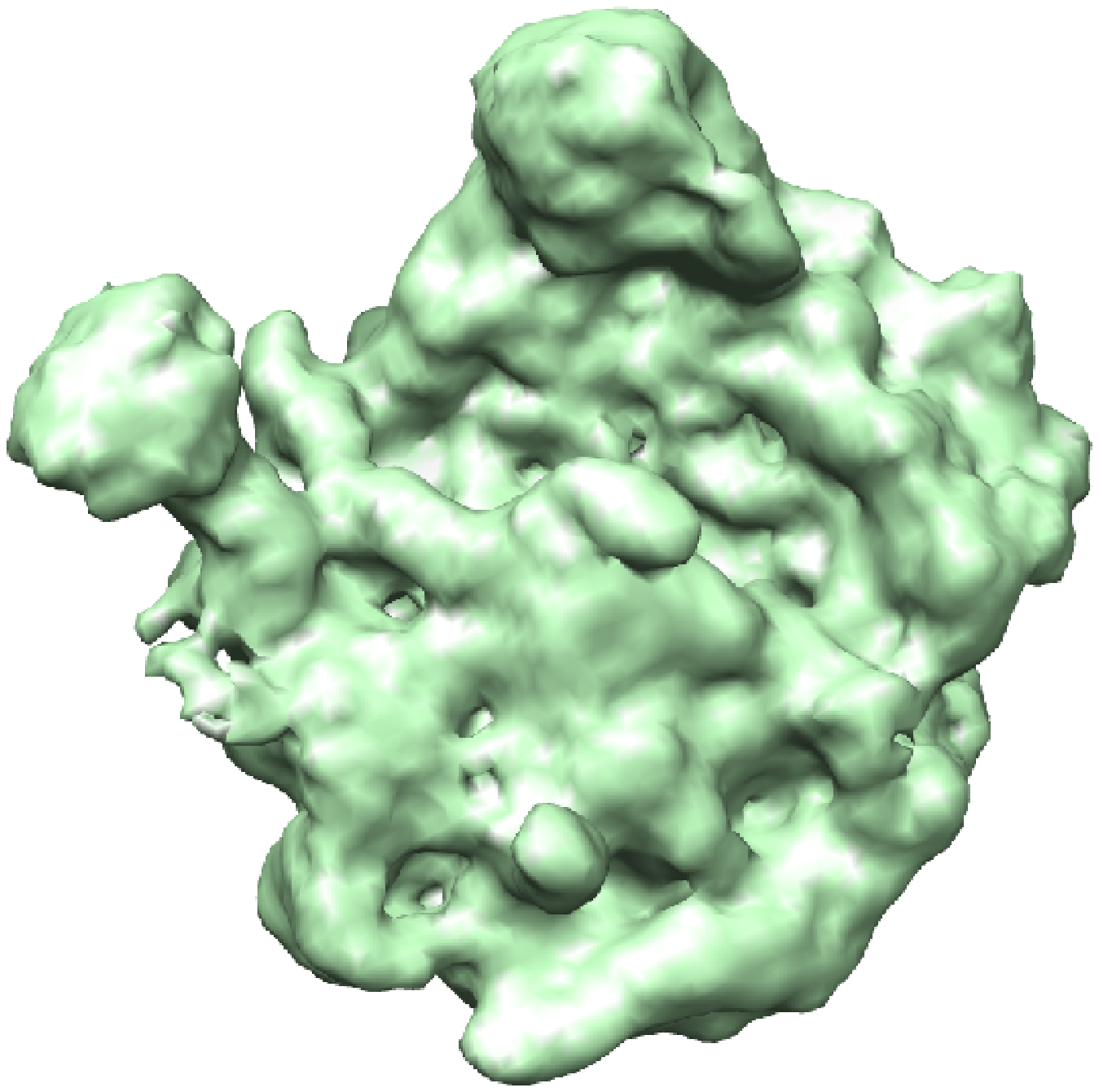} %
\label{fig:FredData_vol2}
}
\end{center}
\caption[50S ribosomal subunit: ab initio reconstructed volumes.]{\textit{Ab initio} reconstructions of 50S ribosomal subunit from two independent data sets. (a) Snapshot of reconstructed volume 1. (b) Snapshot of reconstructed volume 2. }
\label{fig:FredData_volumes}
\end{figure}
\begin{figure}
\begin{center}
\includegraphics[width=0.7\textwidth]{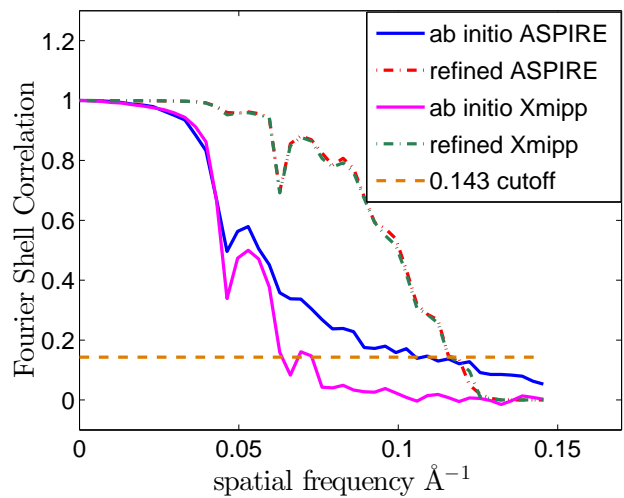}%
\end{center}
\caption{Fourier shell correlation curves for ab initio models and refined models. With $0.143$ cutoff criterion, the resolution is $9.75\text{\AA}$ for ASPIRE ab initio model (blue) and $15.91\text{\AA}$ for Xmipp ab initio model (magenta). Both refined models achieve $8.64\text{\AA}$ resolution according to gold-standard FSC (green and red dot-dash lines).}
\label{fig:50S_fsc}
\end{figure}

We split the data set randomly into two groups of size $13,560$ to generate class averages and reconstructions separately. Each image was identified with 50 nearest neighbors (including reflection) and aligned to get class averaged images. We randomly chose $200$ class averages in each group to build the \textit{ab initio} models with the common-lines based method~\citet{SingerSchkolnisky, Wang2013} for orientation determination.  Figure \ref{fig:FredData} shows 5 arbitrarily chosen class averaged images produced by our algorithm. The two volumes (see Figure \ref{fig:FredData_volumes}) are consistent with each other up to $9.75\text{\AA}$ with 0.143 cutoff criterion (see blue line in Figure~\ref{fig:50S_fsc}). We refined the \textit{ab initio} model using Relion 3D auto-refine~\citet{RELIONtutorial}, and it took 20 iterations to converge to the refined resolution $8.64\text{\AA}$ with gold-standard FSC (see red dot-dash line in Figure~\ref{fig:50S_fsc}).

We used Xmipp CL2D to generate class averages for comparison. CL2D computed $256$ class averages for each group and all class averages were used to build \textit{ab initio} models. The resolution for the \textit{ab initio} model is $15.91\text{\AA}$ with $0.143$ cutoff criterion (see magenta line in Figure~\ref{fig:50S_fsc}). The refined model achieves the same resolution as the refined model from ASPIRE (see Figure~\ref{fig:50S_fsc}). The refinement took 19 iterations to converge, one less iteration than was needed for ASPIRE \textit{ab initio} model. In this example, our class averaging method improved the resolution of the \textit{ab initio} model. However the refinement starting from ASPIRE \textit{ab initio} model did not converge faster than the refinement starting from Xmipp \textit{ab initio} model.  

\subsection{Experimental data: IP$_3$R1}
\begin{figure}
\begin{center}
\includegraphics[width=0.8\textwidth]{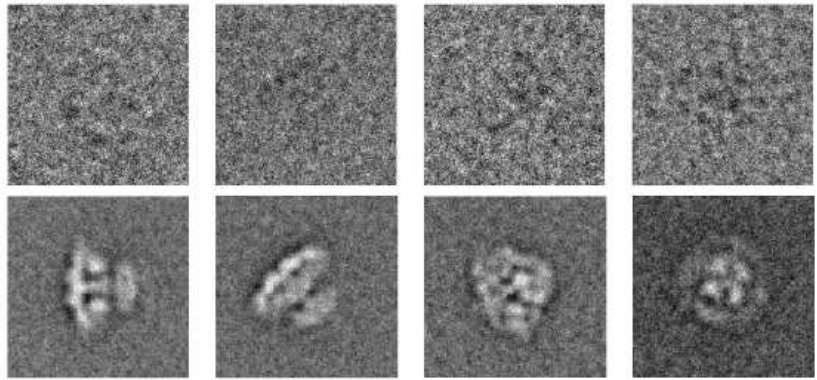} 
\end{center}
\caption{Top row: Samples of experimental images for IP$_3$R1. Bottom row: Class averages obtained by averaging the raw images of the top row with their $50$ aligned nearest neighbors. Courtesy of Dr. Irina Serysheva.}
\label{fig:IP3}
\end{figure}

A set of Inositol $1,4,5$-triphosphate receptor 1 (IP$_3$R1) particle images were provided by Dr. Irina Serysheva. The protein has fourfold symmetry. We are able to generate class averages (the bottom row of Figure~\ref{fig:IP3}) from the original data set (the top row of Figure~\ref{fig:IP3}), which contains $37,382$ images of size $256 \times 256$ pixels. We refer the readers to~\citet{Ludtke2011} for the details of the data set. The experiment shows that our 2D class averaging method, especially the vector diffusion maps classification, also works for particles with non-trivial point group symmetries. The common-lines based \textit{ab initio} orientation determination procedures~\citet{SingerSchkolnisky, Wang2013} have yet to be modified for particles with non-trivial point group symmetry, therefore, we did not attempt to reconstruct the 3D model for this data set. 

\section{Summary and Discussion}
Vitreous-ice-embedded biological macromolecules show a great randomness in orientation. This randomness is exactly what is desired for obtaining high quality $3$D reconstructions. However the variety of viewing angles poses a problem for methods that attempt to rotationally align all images since it is mathematically impossible to bring all images to global alignment. This means that in practice, the distance computed from allegedly globally aligned images is not a rotationally invariant distance. 

In this paper, we introduced a new 2D class averaging procedure. The algorithm has three major components: Fourier-Bessel steerable PCA for image compression and de-noising, bispectrum-like rotational invariant features for classification, and Vector Diffusion Maps for more robust nearest neighbor search and rotational alignment. 

Fourier-Bessel steerable PCA is a fast and accurate procedure for computing the eigen-images of a set of 2D images and their in-plane rotated copies. It is a viable alternative to MSA for compressing and de-noising of the raw 2D images. We demonstrated that this image de-noising method improves the classification results in RFA based classification, MSA/MRA classification, EMAN2 and Xmipp. 

Our rotationally invariant representation of images is based on the bispectrum of their expansion coefficients in the steerable basis. Although the resulting invariant feature vectors are of very high dimensionality, we are able to efficiently project them into a lower dimensional space that captures most variability. Alignment parameters are searched only for nearest neighbors. Reversing the order of alignment and classification leads to a significantly faster viewing angle classification. The algorithm scales almost linearly with the number of images by using a randomized algorithm for nearest neighbor search.

For low SNR, the method that uses direct normalized cross-correlation of the rotationally invariant feature vectors can have many misidentified neighbors. For such situations, Vector Diffusion Maps, a classification method which takes into account the consistency of in-plane rotational transformations between images within the neighborhood, is used to boost the initial viewing angle classification. The eigenvectors of the VDM matrix contain the information of in-plane rotation for nearest neighbor pairs and lead to a much faster and more accurate estimation of the rotational alignments.
  
Through both simulated and experimental data sets, we demonstrated that the new 2D class averaging procedure proposed in this paper is not only fast, but also very robust to noise compared with the commonly used class averaging methods in the field, such as those implemented in SPIDER, IMAGIC, EMAN2, Relion, and Xmipp. The \textit{ab initio} models we built from the experimental data sets are of high resolution and they need fewer iterations of refinement to reach convergence. The methods presented in this paper are also applicable for molecules with non-trivial point group symmetries. The 2D class averaging method described in this paper is freely available as part of our ASPIRE toolbox.

\section*{Acknowledgements}
We would like to thank Yoel Shkolnisky for providing us with his code for the randomized algorithm for PCA~\citet{Halko2011}; Lanhui Wang for sharing her code for orientation estimation using common-lines and for general discussions. We would like to thank Hideki Shigematsu for running the experiments on IMAGIC and Hstau Liao for his help on the 70S ribosome data set. We also thank Sjors Scheres, Steven Ludke, Ignacio Perez, and Carlos Sorzano for helping us with the commonly used cryo-EM software packages. We are indebted to Joachim Frank, Fred Sigworth, Marin van Heel, and Irina Serysheva for providing us with the experimental data sets and for many useful discussions. Parts of this work have appeared in Z. Zhao's PhD dissertation at Princeton University. The project described was supported by Award Number R01GM090200 from the NIGMS, by Award Number FA9550-12-1-0317 and FA9550-13-1-0076 from AFOSR, and by Award Number LTR DTD 06-05-2012 from the Simons Foundation. The content is solely the responsibility of the authors and does not necessarily represent the official views of the National Institute of General Medical Sciences or the National Institutes of Health.

\bibliographystyle{elsarticle-num}
\bibliography{CryoClassAvg}

\end{document}